\documentclass[aps,twocolumn]{revtex4}%
\usepackage{amsfonts}
\usepackage{amsmath}
\usepackage{amssymb}
\usepackage{graphicx}
\usepackage{color}%
\providecommand{\U}[1]{\protect\rule{.1in}{.1in}}
\begin{document}
\title{Interface Between Topological and Superconducting Qubits}
\date{\today}
\author{Liang Jiang,$^{1}$ Charles L. Kane,$^{2}$ John Preskill$^{1}$}
\affiliation{$^{1}$ Institute for Quantum Information, California Institute of Technology,
Pasadena, CA 91125, USA}
\affiliation{$^{2}$ Department of Physics and Astronomy, University of Pennsylvania,
Philadelphia, Pennsylvania 19104, USA}


\begin{abstract}
We propose and analyze an interface between a topological qubit and a
superconducting flux qubit. In our scheme, the interaction between Majorana
fermions in a topological insulator is coherently controlled by a
superconducting phase that depends on the quantum state of the flux qubit. A
controlled phase gate, achieved by pulsing this interaction on and off, can
transfer quantum information between the topological qubit and the
superconducting qubit.

\end{abstract}
\maketitle

\paragraph*{Introduction}

Topologically ordered systems are intrinsically robust against local sources
of decoherence, and therefore hold promise for quantum information processing.
There have been many intriguing proposals for topological qubits, using spin
lattice systems \cite{Kitaev03}, p+ip superconductors \cite{Read00}, and
fractional quantum Hall states with filling factor 5/2 \cite{Nayak08}. The
recently discovered topological insulators \cite{Hasan10} can also support
topologically protected qubits \cite{FuL08}. Meanwhile, conventional systems
for quantum information processing (\emph{e.g.}, ions, spins, photon
polarizations, superconducting qubits) are steadily progressing; recent
developments include high fidelity operations using ions \cite{Blatt08} and
superconducting qubits \cite{Clarke08}, long-distance entanglement generation
using single photons \cite{Moehring07,Togan10}, and extremely long coherence
times using nuclear spins \cite{Dutt07}.

Interfaces between topological and conventional quantum systems have also been
considered recently \cite{Hassler10,Sau10c}. Hybrid systems
\cite{JBLZ08,Aguado08} may allow us to combine the advantages of conventional
qubits (high fidelity readout, universal gates, distributed quantum
communication and computation) with those of topological qubits (robust
quantum storage, protected gates). In this paper, we propose and analyze an
interface between a topological qubit based on Majorana fermions (MFs) at the
surface of a topological insulator (TI) \cite{FuL08} and a conventional
superconducting (SC) flux qubit based on a Josephson junction device
\cite{Mooij99}. The flux qubit has two basis states, for which the SC phase of
a particular SC island has two possible values. In our scheme, this SC phase
coherently controls the interaction between two MFs on the surface of the TI.
This coupling between the MFs and the flux qubit provides a coherent interface
between a topological and conventional quantum system, enabling exchange of
quantum information between the two systems.

\paragraph*{Topological Qubit}

The topological qubit can be encoded with four Majorana fermion operators
$\left\{  \gamma_{i}\right\}  _{i=1,2,3,4}$, which satisfy the Majorana
property $\gamma_{i}^{\dag}=\gamma_{i}$ and fermionic anti-commutation
relation $\left\{  \gamma_{i},\gamma_{j}\right\}  =\delta_{ij}$. A Dirac
fermion operator can be constructed from a pair of MFs $\Gamma_{ij}^{\dag
}=\left(  \gamma_{i}-i\gamma_{j}\right)  /\sqrt{2}$, defining a two
dimensional Hilbert space labeled by $n_{ij}=\Gamma_{ij}^{\dag}\Gamma
_{ij}=0,1$. The two basis states for the topological qubit, each with an even
number of Dirac fermions, are $\left\vert 0\right\rangle _{\mathrm{topo}%
}=\left\vert 0_{12}0_{34}\right\rangle $ and $\left\vert 1\right\rangle
_{\mathrm{topo}}=\left\vert 1_{12}1_{34}\right\rangle $.

\begin{figure}[t]
\begin{center}
\includegraphics[width=7.3cm]{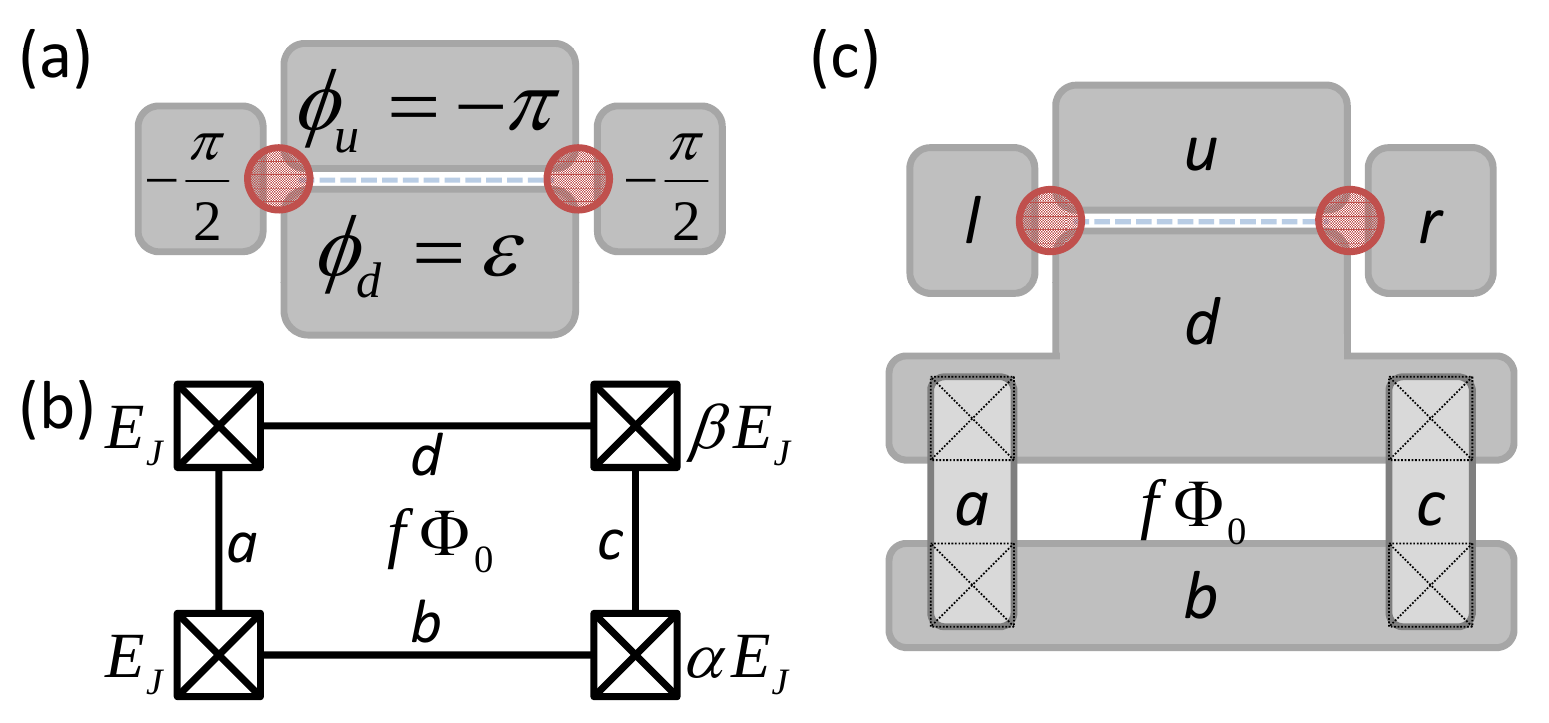}
\end{center}
\caption[fig:TopoFluxQubits]{(color online). On the surface of TI, patterned
SC islands can form (a) STIS quantum wire, (b) flux qubit, and (c) hybrid
system of topological and flux qubits. (a) Two MFs (red dots) are localized at
two SC tri-junctions, connected by an STIS quantum wire (dashed blue line).
The coupling between the MFs is controlled by the SC phases $\phi
_{d}=\varepsilon$ and $\phi_{u}=-\pi$. (b) A flux qubit consists of four JJs
connecting four SC islands (\emph{a,b,c,d}) in series, enclosing an external
magnetic flux $f\Phi_{0}$. (c) The hybrid system consists of an STIS wire and
a flux qubit. The STIS wire (between islands \emph{d} and \emph{u}) couples
the MFs, with coupling strength controlled by the flux qubit. The SC phase
$\phi_{c}$ can be tuned by a phase-controller (not shown), and $\phi_{d}%
=\phi_{c}\pm\theta_{4}^{\ast}$ with the choice of $\pm$ sign depending on the
state of the flux qubit.}%
\label{fig:TopoFluxQubits}%
\end{figure}

The MFs can be created on the surface of a TI patterned with s-wave
superconductors \cite{FuL08}.
Due to the proximity effect \cite{Tinkham96}, Cooper pairs can tunnel into the
TI; hence the effective Hamiltonian describing the surface includes a pairing
term, which has the form $V=\Delta_{0}e^{i\phi}\psi_{\uparrow}^{\dag}%
\psi_{\downarrow}^{\dag}+h.c.$ (where $\psi_{\uparrow}^{\dag}$, $\psi
_{\downarrow}^{\dag}$ are electron operators), assuming that the chemical
potential is close to the Dirac point \cite{Stanescu10}.
Here $\phi$ is the SC phase of the island. Each MF is localized at an SC
vortex that is created by an SC tri-junction (i.e., three separated SC islands
meeting at a point, see Fig.~\ref{fig:TopoFluxQubits}(a)). The MFs can
interact via a superconductor-TI-superconductor (STIS) wire
(Fig.~\ref{fig:TopoFluxQubits}(a)) that separates the SC islands $d$ and $u$
with $\phi_{d}=\varepsilon$ and $\phi_{u}=-\pi$, respectively. For a narrow
STIS wire with width $W\ll v_{F}/\Delta_{0}$, the effective\ Hamiltonian is%
\begin{equation}
H^{\mathrm{STIS}}=-iv_{F}\tau^{x}\partial_{x}+\delta_{\varepsilon}\tau
^{z}\mathrm{,\ }%
\end{equation}
where $v_{F}$ is the effective fermi velocity, $\delta_{\varepsilon}%
=\Delta_{0}\cos\left(  \phi_{d}-\phi_{u}\right)  /2=-\Delta_{0}\sin
\varepsilon/2$, and $\tau^{x,z}$ are Pauli matrices acting on the wire's two
zero energy modes \cite{FuL08}. As shown in Fig.~\ref{fig:TopoFluxQubits}(a),
the STIS wire connects two localized MFs (indicated by two red dots at the
tri-junctions) separated by distance $L$; these are two of the four MFs
comprising the topological qubit. The coupling between the MFs (denoted as
$\gamma_{1}$ and $\gamma_{2}$) via the STIS wire can be characterized by the
Hamiltonian $\tilde{H}_{12}^{\mathrm{MF}}=iE\left(  \varepsilon\right)
\gamma_{1}\gamma_{2}/2$, with an induced energy splitting $E\left(
\varepsilon\right)  $ depending on the SC phase $\varepsilon$. The effective
Hamiltonian for the topological qubit is%
\begin{equation}
H_{12}^{\mathrm{MF}}=-\frac{E\left(  \varepsilon\right)  }{2}\mathbf{Z}%
_{\mathrm{topo}}. \label{eq:MF}%
\end{equation}
where $Z_{\mathrm{topo}}=\left(  \left\vert 0\right\rangle \left\langle
0\right\vert -\left\vert 1\right\rangle \left\langle 1\right\vert \right)
_{\mathrm{topo}}$.

\begin{figure}[t]
\begin{center}
\includegraphics[width=7.0cm]{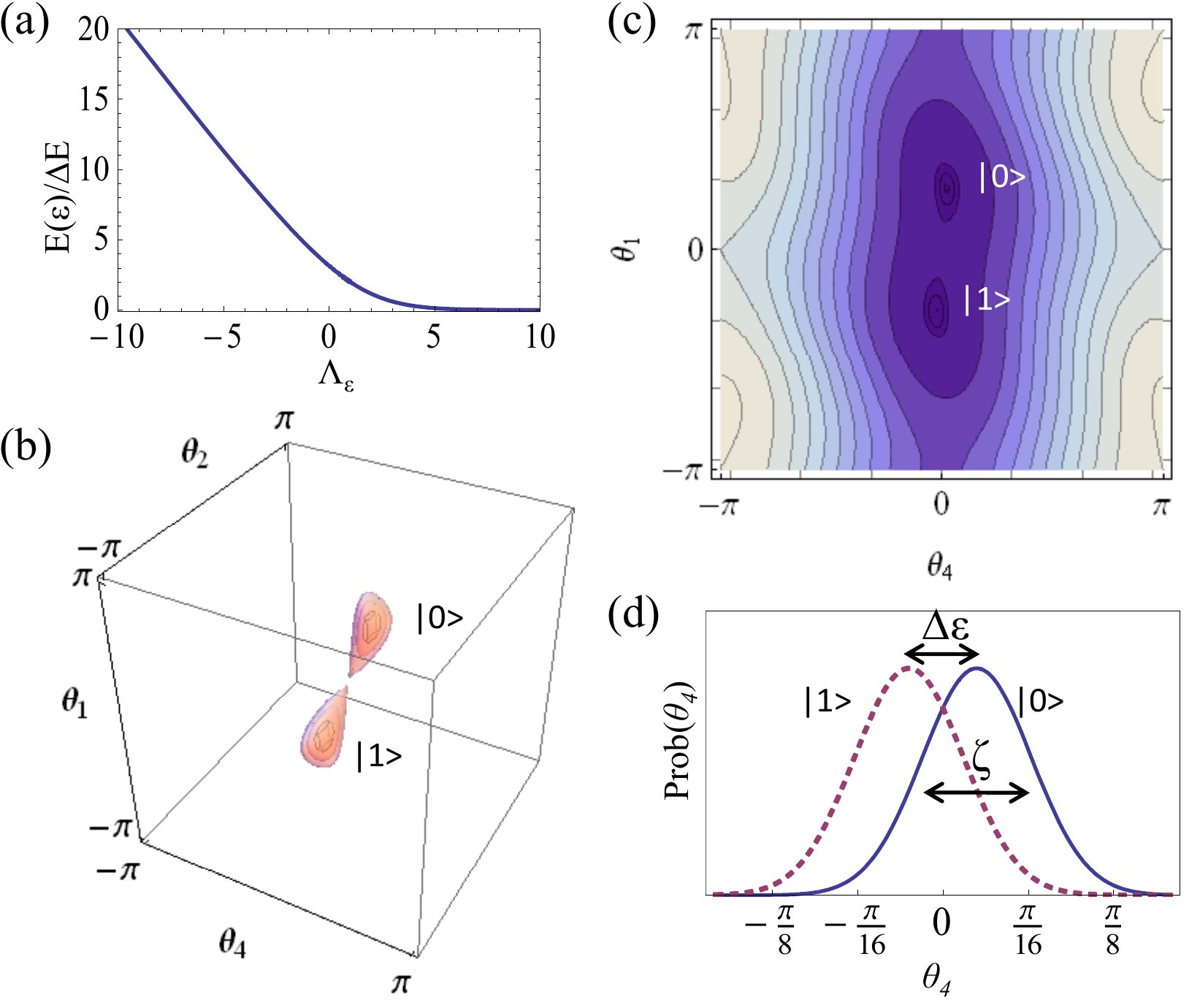}
\end{center}
\caption[fig:Potential]{(color online). (a) The energy splitting $E\left(
\varepsilon\right)  $ (in units of $\Delta E=v_{F}/L$) as a function of
$\Lambda_{\varepsilon}=\frac{\Delta_{0}L}{v_{F}}\sin\varepsilon/2$. (b) A
contour plot of potential energy $U$ as a function of $\left\{  \theta
_{1},\theta_{2},\theta_{4}\right\}  $ with $\theta_{3}=\pi-\theta_{1}%
-\theta_{2}-\theta_{4}$. There are two potential minima associated with flux
qubit states $\left\vert 0\right\rangle $ and $\left\vert 1\right\rangle $.
(c) A contour plot of $U$ as a function of $\left\{  \theta_{1},\theta
_{4}\right\}  $ with $\theta_{1}=\theta_{2}$ and $\theta_{3}=\pi-2\theta
_{1}-\theta_{4}$. (d) Marginal probability distributions of $\theta_{4}$
associated with states $\left\vert 0\right\rangle $ (blue solid line) and
$\left\vert 1\right\rangle $ (red dashed line). The parameters are
$E_{J}/E_{C}=80$ and $\left\{  E_{J,i}/E_{J}\right\}  _{i=1,2,3,4}=\left\{
1,1,\alpha=0.8,\beta=10\right\}  $.}%
\label{fig:Potential}%
\end{figure}

In Fig.~\ref{fig:Potential}(a), we plot $E\left(  \varepsilon\right)  $ as a
function of a dimensionless parameter $\Lambda_{\varepsilon}\equiv\frac
{\Delta_{0}L}{v_{F}}\sin\frac{\varepsilon}{2}$. For $\Lambda_{\varepsilon}%
\gg1$ and $0<\varepsilon<\pi/2$ \cite{FuL08}, the energy splitting $E\left(
\varepsilon\right)  \approx2\left\vert \delta_{\varepsilon}\right\vert
e^{-\Lambda_{\varepsilon}}\sim0$ is negligibly small for localized MFs at the
end of the wire, as the wavefunctions are proportional to $e^{-\Lambda
_{\varepsilon}x/L}$ and $e^{-\Lambda_{\varepsilon}\left(  L-x\right)  /L}$. On
the other hand, for $\Lambda_{\varepsilon}\lesssim1$, the two MFs are
delocalized and $E\left(  \varepsilon\right)  $ becomes sensitive to
$\varepsilon$. We emphasize that $E\left(  \varepsilon\right)  $ is a
non-linear function of $\varepsilon$ \cite{JKP10}, which enables us to switch
the coupling on and off.

\paragraph*{Flux Qubit}

The SC island $d$ can also be part of an SC flux qubit
(Fig.~\ref{fig:TopoFluxQubits}(b)), with $\phi_{d}=\varepsilon=\varepsilon
^{0}$ or $\varepsilon^{1}$ depending on whether the state of the flux qubit is
$\left\vert 0\right\rangle _{\mathrm{flux}}$ or $\left\vert 1\right\rangle
_{\mathrm{flux}}$ as shown in Fig.~\ref{fig:Potential}(b,c). Therefore, the
Hamiltonian $H_{12}^{\mathrm{MF}}$ couples the flux qubit and the topological
qubit. Assuming a small \textrm{\emph{phase separation} }$\Delta
\varepsilon\equiv\varepsilon^{0}-\varepsilon^{1}\ll\pi/2$, we can switch off
the coupling $H_{12}^{\mathrm{MF}}$ by tuning $\varepsilon^{0,1}$ to satisfy
$v_{F}/L\Delta_{0}\ll\varepsilon^{0,1}<\pi/2$ \cite{FuL08}, so that the MFs
are localized and uncoupled with negligible energy splitting $E\left(
\varepsilon^{0}\right)  \approx E\left(  \varepsilon^{1}\right)  \sim0$. We
can also switch on the coupling $H_{12}^{\mathrm{MF}}$ by adiabatically
ramping to the parameter regime $\varepsilon^{0,1}\lesssim v_{F}/L\Delta_{0}$
to induce a nonnegligible $\left\vert E\left(  \varepsilon^{0}\right)
-E\left(  \varepsilon^{1}\right)  \right\vert \sim\Delta_{0}\Delta\varepsilon
$. Because flux qubit designs with three Josephson junctions (JJs)
\cite{Mooij99,Orlando99} are not amenable to achieving a small phase
separation $\Delta\varepsilon\ll\pi/2$ (\textrm{Appendix}), we are motivated to
modify the design of the flux qubit by adding more JJs.

As shown in Fig.~\ref{fig:TopoFluxQubits}(b), our proposed flux qubit consists
of a loop of four Josephson junctions in series that encloses an applied
magnetic flux $f\Phi_{0}$ ($f\approx1/2$ and $\Phi_{0}=h/2e$ is the SC flux
quantum). The Hamiltonian for the flux qubit is%
\begin{equation}
H^{\mathrm{flux}}=T+U,
\end{equation}
with Josephson potential energy $U=\sum_{i=1,2,3,4}E_{J,i}\left(  1-\cos
\theta_{i}\right)  $, and capacitive charging energy $T=\frac{1}{2}%
\sum_{i=1,2,3,4}C_{i}V_{i}^{2}$. For the $i$-th JJ, $E_{J,i}$ is the Josephson
coupling energy, $\theta_{i}$ is the gauge-invariant phase difference, $C_{i}$
is the capacitance, and $V_{i}$ is the voltage across the junction
\cite{Mooij99,Orlando99}. In addition, there are relations satisfied by the
phase accumulation around the loop $\sum_{i}\theta_{i}+2f\pi\equiv0\left(
\operatorname{mod}2\pi\right)  $ and the voltage across each junction
$V_{i}=\left(  \frac{\Phi_{0}}{2\pi}\right)  \dot{\theta}_{i}$
\cite{Tinkham96}. The parameters are chosen as follows: the first two JJs have
equal Josephson coupling energy $E_{J,1}=E_{J,2}=E_{J}$, the third JJ has
$E_{J,3}=\alpha E_{J}$ with $0.5<\alpha<1$, and the fourth JJ has
$E_{J,4}=\beta E_{J}$ with $\beta\gg1$. For JJs with the same thickness but
different junction area $\left\{  A_{i}\right\}  $, $E_{J,i}\propto A_{i}$ and
$C_{i}\propto A_{i}$. The charging energies can be defined as $E_{C,1}%
=E_{C,2}=E_{C}=\frac{e^{2}}{2C_{1}}$, $E_{C,3}=\alpha^{-1}E_{C}$ and
$E_{C,4}=\beta^{-1}E_{C}$. For these parameters and $f\approx1/2$, the system
has two stable states with persistent circulating current of opposite sign. We
identify the flux qubit basis states with the two potential minima $\left\vert
0\right\rangle _{\mathrm{flux}}=\left\vert \left\{  \theta_{i}^{\ast}\right\}
\right\rangle $ and $\left\vert 1\right\rangle _{\mathrm{flux}}=\left\vert
\left\{  -\theta_{i}^{\ast}\right\}  \right\rangle $ (modulo $2\pi$), as
illustrated in Fig.~\ref{fig:Potential}(b,c).

When $\beta\rightarrow\infty$, we may neglect the fourth junction and this
system reduces to the previous flux qubit design with three JJs
\cite{Mooij99,Orlando99}. For $\beta\gg1$, there is a small phase difference
across the fourth JJ (\textrm{Appendix}), $\theta_{4}=\pm\theta_{4}^{\ast}\approx
\pm\frac{\sqrt{4\alpha^{2}-1}}{2\alpha}\frac{1}{\beta}$, where the choice of
$\pm$ sign depends on the direction of the circulating current. We may write
$\theta_{4}=\mathbf{Z}_{\mathrm{flux}}\theta_{4}^{\ast}$, with $\mathbf{Z}%
_{\mathrm{flux}}=\left(  \left\vert 0\right\rangle \left\langle 0\right\vert
-\left\vert 1\right\rangle \left\langle 1\right\vert \right)  _{\mathrm{flux}%
}$. The fourth JJ connects SC islands $c$ and $d$, and if we fix $\phi_{c}$
\textrm{relative to }$\phi_{u}$ with a phase-controller \footnote{An SC
phase-controller can fix the phase difference between two SC islands. It can
be implemented by controlling either the external flux or the current
(\textrm{Appendix}).}, then $\phi_{d}$ will be $\varepsilon^{0}=\phi_{c}+\theta
_{4}^{\ast}$ or $\varepsilon^{1}=\phi_{c}-\theta_{4}^{\ast}$ depending on the
state of the flux qubit. The separation
\begin{equation}
\Delta\varepsilon\approx\frac{\sqrt{4\alpha^{2}-1}}{\alpha}\frac{1}{\beta}%
\end{equation}
between the two possible values of $\phi_{d}$ becomes small, as we desired,
when $\beta$ is large.

\begin{figure}[t]
\begin{center}
\includegraphics[width=5.5cm]{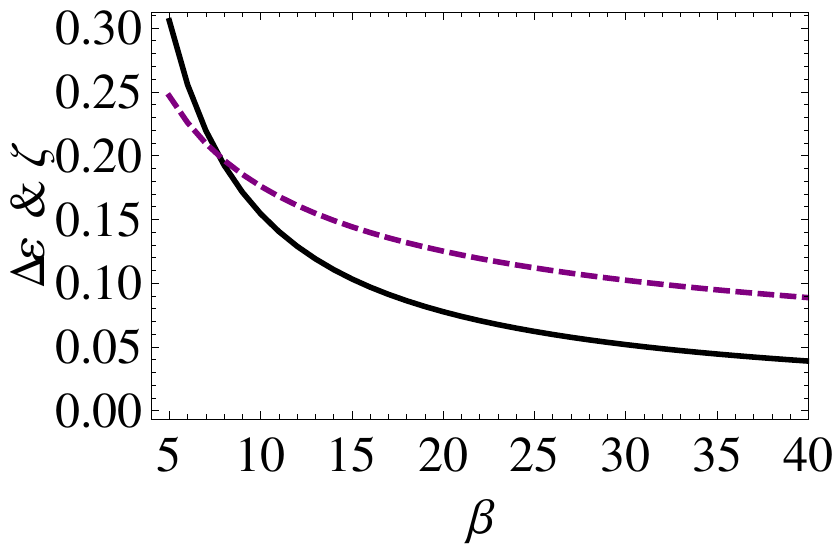}
\end{center}
\caption[fig:BetaDependence]{(color online). Comparison between the phase
separation $\Delta\varepsilon\propto\beta^{-1}$ (dark solid line) and the
magnitude of quantum fluctuations $\zeta\propto\beta^{-1/2}$ (purple dashed
line), assuming $E_{J}/E_{C}=80$.}%
\label{fig:BetaDependence}%
\end{figure}

Aside from this small phase separation, there are also \emph{quantum
fluctuations} in $\theta_{4}$ due to the finite capacitance. Near its minimum
at $\pm\left\{  \theta_{i}^{\ast}\right\}  $, the potential energy is
approximately quadratic; therefore, for $\beta\gg1$, the dynamics of
$\theta_{4}$ can be well described by a harmonic oscillator (HO) Hamiltonian%
\begin{equation}
H^{\mathrm{HO}}=\frac{p_{\theta_{4}}^{2}}{2M_{4}}+\frac{E_{J,4}}{2}\left(
\theta_{4}-\mathbf{Z}_{\mathrm{flux}}\theta_{4}^{\ast}\right)  ^{2},
\label{eq:Osc}%
\end{equation}
where the effective mass is $M_{4}=\frac{1}{8E_{C,4}}$ and the canonical
momentum $p_{\theta_{4}}$ satisfies $\left[  \theta_{4},p_{\theta_{4}}\right]
=i$ (with $\hbar\equiv1$). We may rewrite $H^{\mathrm{HO}}=\left(  a^{\dag
}a+1/2\right)  \omega$ and $\theta_{4}=\mathbf{Z}_{\mathrm{flux}}\theta
_{4}^{\ast}+\zeta\left(  a^{\dag}+a\right)  /\sqrt{2}$, where the oscillator
frequency is $\omega=\sqrt{8E_{J}E_{C}}$ and the magnitude of quantum
fluctuations is $\zeta=\left(  \frac{8E_{C}}{E_{J}}\right)  ^{1/4}\beta
^{-1/2}$. Fig.~\ref{fig:Potential}(d) shows the probability distribution
functions $p_{0/1}\left(  \theta_{4}\right)  \approx\frac{1}{\zeta\sqrt{\pi}%
}e^{-\left(  \theta_{4}\mp\theta_{4}^{\ast}\right)  ^{2}/\zeta^{2}}$
associated with $\left\vert 0\right\rangle _{f}$ and $\left\vert
1\right\rangle _{f}$. The magnitude of the quantum fluctuations $\zeta$ is
comparable to the phase separation $\Delta\varepsilon$; indeed $\zeta
\propto\beta^{-1/2}$ may even dominate the phase separation $\Delta
\varepsilon\propto\beta^{-1}$ for large $\beta$ (Fig.~\ref{fig:BetaDependence}%
). \footnote{Despite the large quantum fluctuations in $\theta_{4}$
($\zeta>\Delta\varepsilon$), the two quantum states of the flux qubit are well
localized because the two potential minima are widely separated in the
$\theta_{1}$ direction (see Fig.~\ref{fig:Potential}(c)).}
Therefore, we should consider both the phase separation and the quantum fluctuations.

\paragraph*{Hybrid System}

The Hamiltonian for the hybrid system of topological and flux qubits
(Fig.~\ref{fig:TopoFluxQubits}(c)) is:%
\begin{equation}
H=H^{\mathrm{HO}}+H_{12}^{\mathrm{MF}}=\left(  a^{\dag}a+1/2\right)
\omega-\frac{1}{2}E\left(  \varepsilon\right)  \mathbf{Z}_{\mathrm{topo}}%
\end{equation}
where $\varepsilon=\phi_{c}+\theta_{4}=\phi_{c}+\mathbf{Z}_{\mathrm{flux}%
}\theta_{4}^{\ast}+\zeta\left(  a^{\dag}+a\right)  /\sqrt{2}$. In both flux
qubit basis states, the oscillator is in its ground state with $\left\langle
a^{\dag}a\right\rangle =0$. To first order in the small parameter
$\delta\equiv\left.  \frac{\zeta}{\omega}\frac{dE\left(  \phi\right)  }{d\phi
}\right\vert _{\phi=\phi_{c}}\ll1$, the Hamiltonian becomes%
\[
H=H^{\mathrm{HO}}-\frac{1}{2}\left(  \left\langle E_{0}\right\rangle
\left\vert 0\right\rangle \left\langle 0\right\vert +\left\langle
E_{1}\right\rangle \left\vert 1\right\rangle \left\langle 1\right\vert
\right)  _{\mathrm{flux}}\otimes\mathbf{Z}_{\mathrm{topo}}+O\left(  \delta
^{2}\right)
\]
where $\left\langle E_{0/1}\right\rangle \equiv\int d\theta_{4}E\left(
\phi_{c}+\theta_{4}\right)  p_{0/1}\left(  \theta_{4}\right)  $.

Up to a single-qubit rotation, the effective Hamiltonian coupling the flux and
topological qubits is%
\begin{equation}
H_{I}=\frac{g}{4}~\mathbf{Z}_{\mathrm{flux}}\mathbf{Z}_{\mathrm{topo}}%
\end{equation}
with coupling strength $g=\left\langle E_{1}\right\rangle -\left\langle
E_{0}\right\rangle \approx\left(  E\left(  \varepsilon^{1}\right)  -E\left(
\varepsilon^{0}\right)  \right)  +\frac{1}{4}\left(  E^{\prime\prime}\left(
\varepsilon^{1}\right)  -E^{\prime\prime}\left(  \varepsilon^{0}\right)
\right)  \zeta^{2}+O\left(  \zeta^{3}\right)  .$The first term arises from the
phase separation and the second term from the quantum fluctuations;
corrections higher order in $\zeta\ll1$ are small.

Because the energy splitting function $E\left(  \varepsilon\right)  $ is
highly non-linear, we may tune $\phi_{c}$ to $\phi_{\mathrm{off}}$ such that
$v_{F}/L\Delta_{0}\ll\varepsilon^{0,1}=\phi_{\mathrm{off}}\pm\Delta
\varepsilon/2<\pi/2$ and switch off the coupling $g\approx\Delta_{0}%
\Delta\varepsilon e^{-\left\vert \phi_{\mathrm{off}}\right\vert \Delta
_{0}L/2v_{F}}\sim0$. On the other hand, we may adiabatically ramp $\phi_{c}$
to $\phi_{\mathrm{on}}\lesssim v_{F}/L\Delta_{0}$, which effectively switches
on the coupling $g\approx\Delta_{0}\Delta\varepsilon$. By adiabatically
changing $\phi_{c}$ from $\phi_{\mathrm{off}}$ $\rightarrow$ $\phi
_{\mathrm{on}}$ $\rightarrow$ $\phi_{\mathrm{off}}$ with $\int g\left(
t\right)  dt=\pi$, we can implement the controlled-phase (\textrm{CP}$_{t,f}$)
gate between the topological ($t$) and flux ($f$) qubits. With Hadamard gates
$\mathrm{Had}_{f}$, we can achieve \textrm{CNOT}$_{t,f}=\mathrm{Had}_{f}%
\cdot\mathrm{CP}_{t,f}\cdot\mathrm{Had}_{f}$, which flips the flux qubit
conditioned on $\left\vert 1\right\rangle _{t}$ and can be used for quantum
non-demolition measurement of the topological qubit \cite{Hassler10,Sau10}.
Furthermore, with Hadamard gates $\mathrm{Had}_{t}$ (implemented by exchanging
two MFs \cite{Nayak08,FuL08}), we can achieve the swap operation
$\mathrm{SWAP}_{t,f}=\left(  \mathrm{Had}_{t}\cdot\mathrm{Had}_{f}%
\cdot\mathrm{CP}_{t,f}\right)  ^{3}$. Finally, with \textrm{CP}$_{t,f}$,
$\mathrm{Had}_{t}$ and single-qubit rotations $\mathrm{U}_{f}$, we can achieve
arbitrary unitary transformations for the two-qubit hybrid system of flux and
topological qubits \cite{JBLZ08,Aguado08}.

\paragraph*{Imperfections}

There are four relevant imperfections for the coupled system of flux and
topological qubits \footnote{We assume that flux qubit parameters can be
accurately measured, and hence ignore fabrication uncertainties.}. The first
imperfection is related to the tunneling between $\left\vert 0\right\rangle
_{\mathrm{flux}}$ and $\left\vert 1\right\rangle _{\mathrm{flux}}$ of the flux
qubit, with tunneling rate $t\sim\omega\exp\left(  -\sqrt{E_{J}/E_{C}}\right)
$. The coupling between flux and topological qubits should be strong enough,
$g\gg t$, to suppress the undesired tunneling probability $\eta_{tunnel}%
\approx\left(  t/g\right)  ^{2}$.

The next imperfection comes from undesired excitations of the oscillators.
According to the Hamiltonian $H$ for the hybrid system, the oscillators may be
excited via interaction $E\left(  \phi_{c}+\theta_{4}\right)  =E\left(
\phi_{c}+\mathbf{Z}_{\mathrm{flux}}\theta_{4}^{\ast}\right)  +\frac
{dE}{d\varepsilon}\zeta\frac{\hat{a}_{1}^{\dag}+\hat{a}_{1}}{\sqrt{2}}+\cdots
$. The excitation probability can be estimated as $\eta_{excite}\approx\left(
\frac{\zeta}{2\omega}\frac{dE}{d\varepsilon}\right)  ^{2}$. Since $\left\vert
\frac{dE}{d\varepsilon}\right\vert \lesssim\Delta_{0}$, $\zeta\approx\left(
\frac{8E_{C}}{E_{J}}\right)  ^{1/4}\beta^{-1/2}$, and $\omega=\sqrt
{8E_{J}E_{C}}$, we estimate $\eta_{excite}\lesssim\frac{1}{20\beta}\left(
\frac{\Delta_{0}}{E_{J}}\right)  ^{2}\sqrt{\frac{E_{J}}{E_{C}}}$.

The third imperfection is due to the finite length of the STIS wire, which
limits the fidelity for the topological qubit itself. When we switch off the
coupling between the flux and topological qubits by having $\phi_{c}%
=\phi_{\mathrm{off}}$ and $\Lambda_{\phi_{\mathrm{off}}}\gg1$ for the STIS
wire, there is an exponentially small energy splitting $E\sim\Delta
_{0}e^{-\Lambda_{\phi_{\mathrm{off}}}}$.

The last relevant imperfection is associated with the excitation modes of the
quantum wire, with excitation energy $E^{\prime}\approx v_{F}/L$ \cite{FuL08}.
Occupation of these modes can potentially modify the phase separation of the
flux qubit. Therefore, we need sufficiently low temperature to exponentially
suppress the occupation of these modes by the factor $e^{-E^{\prime}/k_{B}T}$.

\paragraph*{Physical Parameters}

We may choose the following design parameters for the flux qubit: $\alpha
=0.8$, $\beta=10$, $E_{J}/E_{C}=80$, and $E_{J}=200\left(  2\pi\right)  $ GHz.
Both phase separation and quantum fluctuations depend sensitively on $\beta$
(see Fig. \ref{fig:BetaDependence}), with $\Delta\varepsilon\approx0.16$ and
$\zeta\approx\allowbreak0.18$. Meanwhile, the flux qubit has plasma
oscillation frequency $\omega\approx\allowbreak60\left(  2\pi\right)  $ GHz,
energy barrier $\Delta U\approx0.26E_{J}$, tunneling rate $t\approx
1.8\sqrt{E_{J}E_{C}}\exp\left[  -0.7\left(  E_{J}/E_{C}\right)  ^{1/2}\right]
\approx70\left(  2\pi\right)  $ MHz; these parameters only marginally depend
on $\beta$ (\textrm{Appendix}).

For mesoscopic aluminum junctions with critical current density $500$
A/cm$^{2}$, the largest junction ($E_{J,4}=\beta E_{J}$) has an area of about
$1$ $\mu m^{2}$ \cite{Mooij99}. For the topological qubit, it is feasible to
achieve the parameters $\Delta_{0}\sim0.1$meV$\approx25\left(  2\pi\right)
$GHz, $v_{F}\sim10^{5}$m/s, $L\sim5\mu m$, and $T=20$mK. For the interface,
the effective coupling is $g\sim\Delta_{0}\Delta\varepsilon\sim2\left(
2\pi\right)  $GHz. Therefore, we have imperfections $\eta_{tunnel}\sim10^{-3}%
$, $\eta_{excite}\lesssim10^{-3}$, $e^{-\Lambda_{\phi_{\mathrm{off}}}}\approx
e^{-20\left\vert \sin\phi_{\mathrm{off}}/2\right\vert }<10^{-3}$ (assuming
$\phi_{\mathrm{off}}\approx\pi/4$), and $e^{-E^{\prime}/k_{B}T}<10^{-3}$
\footnote{The SQUID circuit that measures the flux qubit is another potential
source of error, as it may introduce an additional plasma mode with low
frequency $\omega^{\prime}$. Assuming $\omega^{\prime}\sim2\left(
2\pi\right)  $GHz \cite{Chiorescu03}), we may choose $g=200(2\pi)$MHz and
$t=10(2\pi)$MHz, so that $\omega^{\prime}\gg g\gg t\gg1/T_{2}$. If the flux
qubit's spin-echo coherence time $T_{2}$ is long ($\approx4\mu$s
\cite{Bertet05}), a 1\% error rate can be achieved.}.

\paragraph*{Phase Qubit}

A similar interface can be constructed to couple the SC phase qubit
\cite{Martinis02,Clarke08} and the topological qubit. A phase qubit is just a
JJ with\ a fixed DC-current source $I$. The phase qubit Hamiltonian is
$H^{\mathrm{phase}}=T+U^{\mathrm{phase}}$, where $T=\frac{1}{2}E_{C}V^{2}$ and
$U^{\mathrm{phase}}=-I\Phi_{0}\phi-I_{0}\Phi_{0}\cos\phi$. The qubit can be
encoded in the two lowest energy states, $\left\vert 0\right\rangle
_{\mathrm{phase}}$ and $\left\vert 1\right\rangle _{\mathrm{phase}}$, with
magnitude of quantum fluctuations $\zeta_{0}$ and $\zeta_{1}$, respectively.
The coupling strength between phase and topological qubits can be estimated as
$g^{\mathrm{phase}}\approx E^{\prime\prime}\left(  \varepsilon\right)  \left(
\zeta_{1}^{2}-\zeta_{0}^{2}\right)  $.

\paragraph*{Notes added}

It was recently proposed to use the Aharonov-Casher (AC) effect for quantum
non-demolition measurement of a topological qubit \cite{Hassler10,Sau10c}.
This proposal, which applies in the parameter regime $\alpha>1$ where the flux
qubit has two possible tunneling pathways, exploits the observation that
whether two tunneling paths interfere destructively or constructively can be
controlled by the state of the topological qubit. In contrast, our proposal,
which applies in the parameter regime $\alpha<1$ where the flux qubit has only
one tunneling pathway, exploits the non-linearity of the energy splitting
$E\left(  \varepsilon\right)  $ to achieve a controlled-phase coupling between
the topological and flux qubits. Recently, the related
work \cite{Bonderson10b} appeared.

\paragraph*{Conclusion}

We have proposed and analyzed a feasible interface between flux and
topological qubits. Our proposal uses a flux qubit design with four JJs, such
that the two basis states of the qubit have a small phase separation
$\Delta\varepsilon$ on a particular superconducting island, enabling us to
adiabatically switch on and off the coupling between the flux and topological
qubits. Such interfaces may enable us to store and retrieve quantum
information using the topological qubit, to repetitively readout the
topological qubit with a conventional qubit, or to switch between conventional
and topological systems for various quantum information processing tasks.

We are especially indebted to Mikhail Lukin for inspiring discussions. We also
thank Anton Akhmerov, Jason Alicea, Erez Berg, David DiVincenzo, Garry
Goldstein, Netanel Lindner and Gil Refael for helpful comments. This work was
supported by the Sherman Fairchild Foundation, by NSF grants DMR-0906175 and
PHY-0803371, by DOE grant DE-FG03-92-ER40701, and by NSA/ARO grant W911NF-09-1-0442.

\bibliographystyle{apsrev}
\bibliography{refAll}

\begin{thebibliography}{25}
\expandafter\ifx\csname natexlab\endcsname\relax\def\natexlab#1{#1}\fi
\expandafter\ifx\csname bibnamefont\endcsname\relax
  \def\bibnamefont#1{#1}\fi
\expandafter\ifx\csname bibfnamefont\endcsname\relax
  \def\bibfnamefont#1{#1}\fi
\expandafter\ifx\csname citenamefont\endcsname\relax
  \def\citenamefont#1{#1}\fi
\expandafter\ifx\csname url\endcsname\relax
  \def\url#1{\texttt{#1}}\fi
\expandafter\ifx\csname urlprefix\endcsname\relax\def\urlprefix{URL }\fi
\providecommand{\bibinfo}[2]{#2}
\providecommand{\eprint}[2][]{\url{#2}}

\bibitem[{\citenamefont{Kitaev}(2003)}]{Kitaev03}
\bibinfo{author}{\bibfnamefont{A.~Y.} \bibnamefont{Kitaev}},
  \bibinfo{journal}{Annals of Physics} \textbf{\bibinfo{volume}{303}},
  \bibinfo{pages}{2} (\bibinfo{year}{2003}).

\bibitem[{\citenamefont{Read and Green}(2000)}]{Read00}
\bibinfo{author}{\bibfnamefont{N.}~\bibnamefont{Read}} \bibnamefont{and}
  \bibinfo{author}{\bibfnamefont{D.}~\bibnamefont{Green}},
  \bibinfo{journal}{Phys. Rev. B} \textbf{\bibinfo{volume}{61}},
  \bibinfo{pages}{10267} (\bibinfo{year}{2000}).

\bibitem[{\citenamefont{Nayak et~al.}(2008)\citenamefont{Nayak, Simon, Stern,
  Freedman, and Das~Sarma}}]{Nayak08}
\bibinfo{author}{\bibfnamefont{C.}~\bibnamefont{Nayak}},
  \bibinfo{author}{\bibfnamefont{S.~H.} \bibnamefont{Simon}},
  \bibinfo{author}{\bibfnamefont{A.}~\bibnamefont{Stern}},
  \bibinfo{author}{\bibfnamefont{M.}~\bibnamefont{Freedman}}, \bibnamefont{and}
  \bibinfo{author}{\bibfnamefont{S.}~\bibnamefont{Das~Sarma}},
  \bibinfo{journal}{Rev. Mod. Phys.} \textbf{\bibinfo{volume}{80}},
  \bibinfo{pages}{1083} (\bibinfo{year}{2008}).

\bibitem[{\citenamefont{Hasan and Kane}(2010)}]{Hasan10}
\bibinfo{author}{\bibfnamefont{M.~Z.} \bibnamefont{Hasan}} \bibnamefont{and}
  \bibinfo{author}{\bibfnamefont{C.~L.} \bibnamefont{Kane}},
  \bibinfo{journal}{Rev. Mod. Phys.} \textbf{\bibinfo{volume}{82}},
  \bibinfo{pages}{3045} (\bibinfo{year}{2010}).

\bibitem[{\citenamefont{Fu and Kane}(2008)}]{FuL08}
\bibinfo{author}{\bibfnamefont{L.}~\bibnamefont{Fu}} \bibnamefont{and}
  \bibinfo{author}{\bibfnamefont{C.~L.} \bibnamefont{Kane}},
  \bibinfo{journal}{Phys. Rev. Lett.} \textbf{\bibinfo{volume}{100}},
  \bibinfo{pages}{096407} (\bibinfo{year}{2008}).

\bibitem[{\citenamefont{Blatt and Wineland}(2008)}]{Blatt08}
\bibinfo{author}{\bibfnamefont{R.}~\bibnamefont{Blatt}} \bibnamefont{and}
  \bibinfo{author}{\bibfnamefont{D.}~\bibnamefont{Wineland}},
  \bibinfo{journal}{Nature (London)} \textbf{\bibinfo{volume}{453}},
  \bibinfo{pages}{1008} (\bibinfo{year}{2008}).

\bibitem[{\citenamefont{Clarke and Wilhelm}(2008)}]{Clarke08}
\bibinfo{author}{\bibfnamefont{J.}~\bibnamefont{Clarke}} \bibnamefont{and}
  \bibinfo{author}{\bibfnamefont{F.~K.} \bibnamefont{Wilhelm}},
  \bibinfo{journal}{Nature (London)} \textbf{\bibinfo{volume}{453}},
  \bibinfo{pages}{1031} (\bibinfo{year}{2008}).

\bibitem[{\citenamefont{Moehring et~al.}(2007)\citenamefont{Moehring, Maunz,
  Olmschenk, Younge, Matsukevich, Duan, and Monroe}}]{Moehring07}
\bibinfo{author}{\bibfnamefont{D.~L.} \bibnamefont{Moehring}},
  \bibinfo{author}{\bibfnamefont{P.}~\bibnamefont{Maunz}},
  \bibinfo{author}{\bibfnamefont{S.}~\bibnamefont{Olmschenk}},
  \bibinfo{author}{\bibfnamefont{K.~C.} \bibnamefont{Younge}},
  \bibinfo{author}{\bibfnamefont{D.~N.} \bibnamefont{Matsukevich}},
  \bibinfo{author}{\bibfnamefont{L.~M.} \bibnamefont{Duan}}, \bibnamefont{and}
  \bibinfo{author}{\bibfnamefont{C.}~\bibnamefont{Monroe}},
  \bibinfo{journal}{Nature (London)} \textbf{\bibinfo{volume}{449}},
  \bibinfo{pages}{68} (\bibinfo{year}{2007}).

\bibitem[{\citenamefont{Togan et~al.}(2010)\citenamefont{Togan, Chu, Trifonov,
  Jiang, Maze, Childress, Dutt, Sorensen, Hemmer, Zibrov et~al.}}]{Togan10}
\bibinfo{author}{\bibfnamefont{E.}~\bibnamefont{Togan}},
  \bibinfo{author}{\bibfnamefont{Y.}~\bibnamefont{Chu}},
  \bibinfo{author}{\bibfnamefont{A.~S.} \bibnamefont{Trifonov}},
  \bibinfo{author}{\bibfnamefont{L.}~\bibnamefont{Jiang}},
  \bibinfo{author}{\bibfnamefont{J.}~\bibnamefont{Maze}},
  \bibinfo{author}{\bibfnamefont{L.}~\bibnamefont{Childress}},
  \bibinfo{author}{\bibfnamefont{M.~V.~G.} \bibnamefont{Dutt}},
  \bibinfo{author}{\bibfnamefont{A.~S.} \bibnamefont{Sorensen}},
  \bibinfo{author}{\bibfnamefont{P.~R.} \bibnamefont{Hemmer}},
  \bibinfo{author}{\bibfnamefont{A.~S.} \bibnamefont{Zibrov}},
  \bibnamefont{et~al.}, \bibinfo{journal}{Nature (London)}
  \textbf{\bibinfo{volume}{466}}, \bibinfo{pages}{730} (\bibinfo{year}{2010}).

\bibitem[{\citenamefont{Dutt et~al.}(2007)\citenamefont{Dutt, Childress, Jiang,
  Togan, Maze, Jelezko, Zibrov, Hemmer, and Lukin}}]{Dutt07}
\bibinfo{author}{\bibfnamefont{M.~V.~G.} \bibnamefont{Dutt}},
  \bibinfo{author}{\bibfnamefont{L.}~\bibnamefont{Childress}},
  \bibinfo{author}{\bibfnamefont{L.}~\bibnamefont{Jiang}},
  \bibinfo{author}{\bibfnamefont{E.}~\bibnamefont{Togan}},
  \bibinfo{author}{\bibfnamefont{J.}~\bibnamefont{Maze}},
  \bibinfo{author}{\bibfnamefont{F.}~\bibnamefont{Jelezko}},
  \bibinfo{author}{\bibfnamefont{A.~S.} \bibnamefont{Zibrov}},
  \bibinfo{author}{\bibfnamefont{P.~R.} \bibnamefont{Hemmer}},
  \bibnamefont{and} \bibinfo{author}{\bibfnamefont{M.~D.} \bibnamefont{Lukin}},
  \bibinfo{journal}{Science} \textbf{\bibinfo{volume}{316}},
  \bibinfo{pages}{1312} (\bibinfo{year}{2007}).

\bibitem[{\citenamefont{Hassler et~al.}(2010)\citenamefont{Hassler, Akhmerov,
  Hou, and Beenakker}}]{Hassler10}
\bibinfo{author}{\bibfnamefont{F.}~\bibnamefont{Hassler}},
  \bibinfo{author}{\bibfnamefont{A.~R.} \bibnamefont{Akhmerov}},
  \bibinfo{author}{\bibfnamefont{C.-Y.} \bibnamefont{Hou}}, \bibnamefont{and}
  \bibinfo{author}{\bibfnamefont{C.~W.~J.} \bibnamefont{Beenakker}},
  \bibinfo{journal}{New J. Phys.} \textbf{\bibinfo{volume}{12}},
  \bibinfo{pages}{125002} (\bibinfo{year}{2010}).

\bibitem[{\citenamefont{Sau et~al.}(2010{\natexlab{a}})\citenamefont{Sau,
  Tewari, and Das~Sarma}}]{Sau10c}
\bibinfo{author}{\bibfnamefont{J.~D.} \bibnamefont{Sau}},
  \bibinfo{author}{\bibfnamefont{S.}~\bibnamefont{Tewari}}, \bibnamefont{and}
  \bibinfo{author}{\bibfnamefont{S.}~\bibnamefont{Das~Sarma}},
  \bibinfo{journal}{Phys. Rev. A} \textbf{\bibinfo{volume}{82}},
  \bibinfo{pages}{052322} (\bibinfo{year}{2010}{\natexlab{a}}).

\bibitem[{\citenamefont{Jiang et~al.}(2008)\citenamefont{Jiang, Brennen,
  Gorshkov, Hammerer, Hafezi, Demler, Lukin, and Zoller}}]{JBLZ08}
\bibinfo{author}{\bibfnamefont{L.}~\bibnamefont{Jiang}},
  \bibinfo{author}{\bibfnamefont{G.~K.} \bibnamefont{Brennen}},
  \bibinfo{author}{\bibfnamefont{A.}~\bibnamefont{Gorshkov}},
  \bibinfo{author}{\bibfnamefont{K.}~\bibnamefont{Hammerer}},
  \bibinfo{author}{\bibfnamefont{M.}~\bibnamefont{Hafezi}},
  \bibinfo{author}{\bibfnamefont{E.}~\bibnamefont{Demler}},
  \bibinfo{author}{\bibfnamefont{M.~D.} \bibnamefont{Lukin}}, \bibnamefont{and}
  \bibinfo{author}{\bibfnamefont{P.}~\bibnamefont{Zoller}},
  \bibinfo{journal}{Nat. Phys.} \textbf{\bibinfo{volume}{4}},
  \bibinfo{pages}{482} (\bibinfo{year}{2008}).

\bibitem[{\citenamefont{Aguado et~al.}(2008)\citenamefont{Aguado, Brennen,
  Verstraete, and Cirac}}]{Aguado08}
\bibinfo{author}{\bibfnamefont{M.}~\bibnamefont{Aguado}},
  \bibinfo{author}{\bibfnamefont{G.~K.} \bibnamefont{Brennen}},
  \bibinfo{author}{\bibfnamefont{F.}~\bibnamefont{Verstraete}},
  \bibnamefont{and} \bibinfo{author}{\bibfnamefont{J.~I.} \bibnamefont{Cirac}},
  \bibinfo{journal}{Phys. Rev. Lett.} \textbf{\bibinfo{volume}{101}},
  \bibinfo{pages}{260501} (\bibinfo{year}{2008}).

\bibitem[{\citenamefont{Mooij et~al.}(1999)\citenamefont{Mooij, Orlando,
  Levitov, Tian, van~der Wal, and Lloyd}}]{Mooij99}
\bibinfo{author}{\bibfnamefont{J.~E.} \bibnamefont{Mooij}},
  \bibinfo{author}{\bibfnamefont{T.~P.} \bibnamefont{Orlando}},
  \bibinfo{author}{\bibfnamefont{L.}~\bibnamefont{Levitov}},
  \bibinfo{author}{\bibfnamefont{L.}~\bibnamefont{Tian}},
  \bibinfo{author}{\bibfnamefont{C.~H.} \bibnamefont{van~der Wal}},
  \bibnamefont{and} \bibinfo{author}{\bibfnamefont{S.}~\bibnamefont{Lloyd}},
  \bibinfo{journal}{Science} \textbf{\bibinfo{volume}{285}},
  \bibinfo{pages}{1036} (\bibinfo{year}{1999}).

\bibitem[{\citenamefont{Tinkham}(1996)}]{Tinkham96}
\bibinfo{author}{\bibfnamefont{M.}~\bibnamefont{Tinkham}},
  \emph{\bibinfo{title}{Introduction to superconductivity}}
  (\bibinfo{publisher}{McGraw Hill}, \bibinfo{address}{New York},
  \bibinfo{year}{1996}), \bibinfo{edition}{2nd} ed.

\bibitem[{\citenamefont{Stanescu et~al.}(2010)\citenamefont{Stanescu, Sau,
  Lutchyn, and Das~Sarma}}]{Stanescu10}
\bibinfo{author}{\bibfnamefont{T.~D.} \bibnamefont{Stanescu}},
  \bibinfo{author}{\bibfnamefont{J.~D.} \bibnamefont{Sau}},
  \bibinfo{author}{\bibfnamefont{R.~M.} \bibnamefont{Lutchyn}},
  \bibnamefont{and}
  \bibinfo{author}{\bibfnamefont{S.}~\bibnamefont{Das~Sarma}},
  \bibinfo{journal}{Phys. Rev. B} \textbf{\bibinfo{volume}{81}},
  \bibinfo{pages}{241310} (\bibinfo{year}{2010}).

\bibitem[{\citenamefont{Jiang et~al.}(2011)\citenamefont{Jiang, Kane, and
  Preskill}}]{JKP10}
\bibinfo{author}{\bibfnamefont{L.}~\bibnamefont{Jiang}},
  \bibinfo{author}{\bibfnamefont{C.~L.} \bibnamefont{Kane}}, \bibnamefont{and}
  \bibinfo{author}{\bibfnamefont{J.}~\bibnamefont{Preskill}},
  \bibinfo{journal}{arXiv:} \textbf{\bibinfo{volume}{1010.5862}}
  (\bibinfo{year}{2011}).

\bibitem[{\citenamefont{Orlando et~al.}(1999)\citenamefont{Orlando, Mooij,
  Tian, van~der Wal, Levitov, Lloyd, and Mazo}}]{Orlando99}
\bibinfo{author}{\bibfnamefont{T.~P.} \bibnamefont{Orlando}},
  \bibinfo{author}{\bibfnamefont{J.~E.} \bibnamefont{Mooij}},
  \bibinfo{author}{\bibfnamefont{L.}~\bibnamefont{Tian}},
  \bibinfo{author}{\bibfnamefont{C.~H.} \bibnamefont{van~der Wal}},
  \bibinfo{author}{\bibfnamefont{L.~S.} \bibnamefont{Levitov}},
  \bibinfo{author}{\bibfnamefont{S.}~\bibnamefont{Lloyd}}, \bibnamefont{and}
  \bibinfo{author}{\bibfnamefont{J.~J.} \bibnamefont{Mazo}},
  \bibinfo{journal}{Phys. Rev. B} \textbf{\bibinfo{volume}{60}},
  \bibinfo{pages}{15398} (\bibinfo{year}{1999}).

\bibitem[{\citenamefont{Sau et~al.}(2010{\natexlab{b}})\citenamefont{Sau,
  Lutchyn, Tewari, and Das~Sarma}}]{Sau10}
\bibinfo{author}{\bibfnamefont{J.~D.} \bibnamefont{Sau}},
  \bibinfo{author}{\bibfnamefont{R.~M.} \bibnamefont{Lutchyn}},
  \bibinfo{author}{\bibfnamefont{S.}~\bibnamefont{Tewari}}, \bibnamefont{and}
  \bibinfo{author}{\bibfnamefont{S.}~\bibnamefont{Das~Sarma}},
  \bibinfo{journal}{Phys. Rev. Lett.} \textbf{\bibinfo{volume}{104}},
  \bibinfo{pages}{040502} (\bibinfo{year}{2010}{\natexlab{b}}).

\bibitem[{\citenamefont{Martinis et~al.}(2002)\citenamefont{Martinis, Nam,
  Aumentado, and Urbina}}]{Martinis02}
\bibinfo{author}{\bibfnamefont{J.~M.} \bibnamefont{Martinis}},
  \bibinfo{author}{\bibfnamefont{S.}~\bibnamefont{Nam}},
  \bibinfo{author}{\bibfnamefont{J.}~\bibnamefont{Aumentado}},
  \bibnamefont{and} \bibinfo{author}{\bibfnamefont{C.}~\bibnamefont{Urbina}},
  \bibinfo{journal}{Phys. Rev. Lett.} \textbf{\bibinfo{volume}{89}},
  \bibinfo{pages}{117901} (\bibinfo{year}{2002}).

\bibitem[{\citenamefont{Bonderson and Lutchyn}(2010)}]{Bonderson10b}
\bibinfo{author}{\bibfnamefont{P.}~\bibnamefont{Bonderson}} \bibnamefont{and}
  \bibinfo{author}{\bibfnamefont{R.~M.} \bibnamefont{Lutchyn}},
  \bibinfo{journal}{Phys. Rev. Lett.} \textbf{\bibinfo{volume}{106}},
  \bibinfo{pages}{130505} (\bibinfo{year}{2010}).

\bibitem[{\citenamefont{Chiorescu et~al.}(2003)\citenamefont{Chiorescu,
  Nakamura, Harmans, and Mooij}}]{Chiorescu03}
\bibinfo{author}{\bibfnamefont{I.}~\bibnamefont{Chiorescu}},
  \bibinfo{author}{\bibfnamefont{Y.}~\bibnamefont{Nakamura}},
  \bibinfo{author}{\bibfnamefont{C.~J. P.~M.} \bibnamefont{Harmans}},
  \bibnamefont{and} \bibinfo{author}{\bibfnamefont{J.~E.} \bibnamefont{Mooij}},
  \bibinfo{journal}{Science} \textbf{\bibinfo{volume}{299}},
  \bibinfo{pages}{1869} (\bibinfo{year}{2003}).

\bibitem[{\citenamefont{Bertet et~al.}(2005)\citenamefont{Bertet, Chiorescu,
  Burkard, Semba, Harmans, DiVincenzo, and Mooij}}]{Bertet05}
\bibinfo{author}{\bibfnamefont{P.}~\bibnamefont{Bertet}},
  \bibinfo{author}{\bibfnamefont{I.}~\bibnamefont{Chiorescu}},
  \bibinfo{author}{\bibfnamefont{G.}~\bibnamefont{Burkard}},
  \bibinfo{author}{\bibfnamefont{K.}~\bibnamefont{Semba}},
  \bibinfo{author}{\bibfnamefont{C.~J. P.~M.} \bibnamefont{Harmans}},
  \bibinfo{author}{\bibfnamefont{D.~P.} \bibnamefont{DiVincenzo}},
  \bibnamefont{and} \bibinfo{author}{\bibfnamefont{J.~E.} \bibnamefont{Mooij}},
  \bibinfo{journal}{Phys. Rev. Lett.} \textbf{\bibinfo{volume}{95}},
  \bibinfo{pages}{257002} (\bibinfo{year}{2005}).

\bibitem[{\citenamefont{Friedman et~al.}(2000)\citenamefont{Friedman, Patel,
  Chen, Tolpygo, and Lukens}}]{Friedman00}
\bibinfo{author}{\bibfnamefont{J.~R.} \bibnamefont{Friedman}},
  \bibinfo{author}{\bibfnamefont{V.}~\bibnamefont{Patel}},
  \bibinfo{author}{\bibfnamefont{W.}~\bibnamefont{Chen}},
  \bibinfo{author}{\bibfnamefont{S.~K.} \bibnamefont{Tolpygo}},
  \bibnamefont{and} \bibinfo{author}{\bibfnamefont{J.~E.}
  \bibnamefont{Lukens}}, \bibinfo{journal}{Nature (London)}
  \textbf{\bibinfo{volume}{406}}, \bibinfo{pages}{43} (\bibinfo{year}{2000}).

\end{thebibliography}


\appendix

\section{Flux Qubits}

We describe our design of the SC flux qubit that prepares an SC island with SC
phase $\varepsilon=\varepsilon^{0}$ or $\varepsilon^{1}$ depending on the flux
qubit state $\left\vert 0\right\rangle _{\mathrm{flux}}$ or $\left\vert
1\right\rangle _{\mathrm{flux}}$. The phase of the SC island can coherently
control the coupling between the two Majorana fermions at the end of the STIS
wire. We would like to have a small phase separation $\Delta\varepsilon
\equiv\varepsilon^{1}-\varepsilon^{0}\ll\pi/2$, so that we can easily switch
off the coupling $H_{12}^{\mathrm{MF}}$ when we do not want to couple the
Majorana fermions (MFs).

The previous design of flux qubit with three JJs \cite{Mooij99,Orlando99},
however, is not amenable to achieving a small phase separation $\Delta
\varepsilon\ll\pi/2$, because of the following reason. The three JJs have
Josephson energy $E_{J,1}=E_{J,2}=E_{J}$ and $E_{J,3}=\alpha E_{J}$
\cite{Mooij99,Orlando99}. The phase difference across the first junction is
$\theta=\pm\cos^{-1}\frac{1}{2\alpha}$. By choosing $\alpha=\eta+1/2$ and
$\eta\ll1$, there is a small phase separation $\Delta\varepsilon=2\left\vert
\theta\right\vert \approx4\eta^{1/2}$. However, the energy barrier for the
tunneling is significantly suppressed $\Delta U=2\alpha-\left(  2-\frac
{1}{2\alpha}\right)  E_{J}\approx4\eta^{2}E_{J}\sim\theta^{4}E_{J}$. The
action associated with the tunneling is $S\approx\theta\sqrt{\Delta U/E_{C}%
}\approx\theta^{3}\sqrt{E_{J}/E_{C}}\propto\theta^{3}E_{J}$, where the last
step uses the property that $E_{C}\propto1/E_{J}$. In order to maintain a
similar tunneling matrix element between the two potential minima, it requires
the unfavorable scaling $E_{J}\propto1/\theta^{3}$. For example, to achieve
$\theta=0.1$, we have to increase the area of the Josephson junction by
$\sim10^{3}$. Practically, it would also be very challenging to have a precise
value of $\alpha=\frac{1}{2}+\frac{\theta^{2}}{4}=0.5025$, which fine tunes
the phase separation $\Delta\varepsilon$. In contrast, the four-junction
design of flux qubit has a favorable scaling of $E_{J,4}=\beta E_{J}%
\propto1/\theta$ and there is no fine-tuned parameter. This motivates us to
redesign the flux qubit with more Josephson junctions.

\begin{figure}[b]
\begin{center}
\includegraphics[width=6cm]{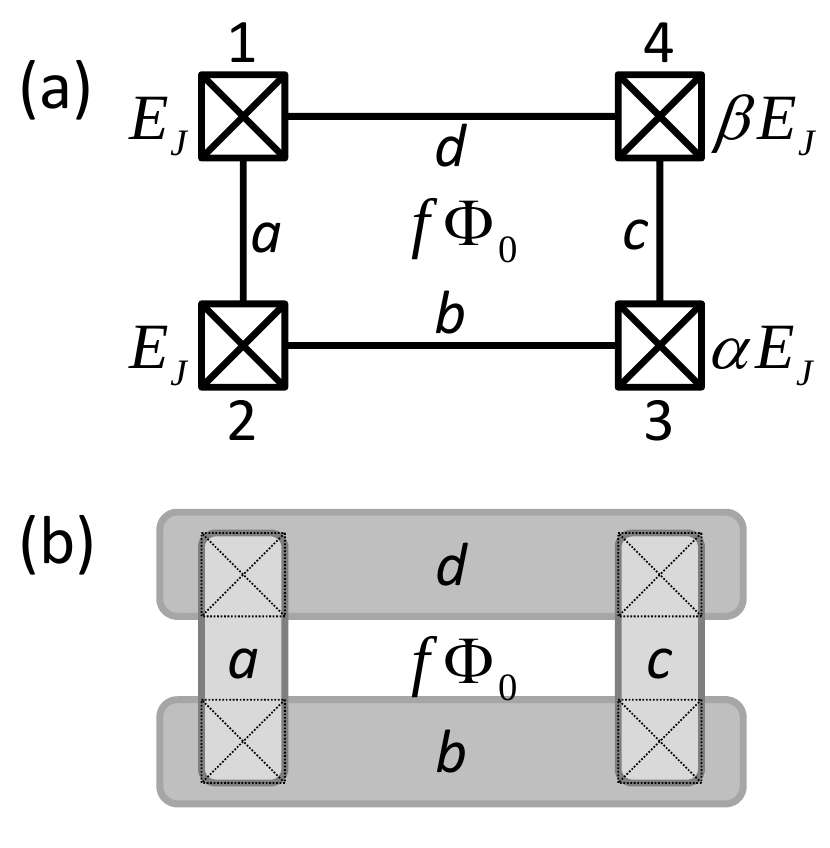}
\end{center}
\caption[fig:FluxQubit]{The design of flux qubit consists of a loop of four
JJs in series that encloses an external magnetic flux $f\Phi_{0}$. Schematic
illustration in terms of (a) JJs and (b) SC islands. }%
\label{fig:FluxQubit}%
\end{figure}

As shown in Fig.~\ref{fig:FluxQubit}, the flux qubit consists of a loop of
four JJs in series that encloses an external magnetic flux $f\Phi_{0}$
($f\approx1/2$ and $\Phi_{0}=h/2e$ is the SC flux quantum). The first two JJs
have equal Josephson coupling energy $E_{J,1}=E_{J,2}=E_{J}$; the third JJ has
coupling energy $E_{J,3}=\alpha E_{J}$, with $0.5<\alpha<1$; the coupling in
the fourth JJ is $E_{J,4}=\beta E_{J}$, with $\beta\gg1$. For JJs with the
same thickness but different junction area $\left\{  A_{j}\right\}  $,
$E_{J,j}\propto A_{j}$ and $C_{j}\propto A_{j}$. The charging energies can be
defined as $E_{C,1}=E_{C,2}=E_{C}=\frac{e^{2}}{2C_{1}}$, $E_{C,3}=\alpha
^{-1}E_{C}$ and $E_{C,4}=\beta^{-1}E_{C}$. Notice that $E_{J,j}E_{C,j}%
=E_{J}E_{C}$ is independent of $j$. The system may have two stable states with
persistent circulating current of opposite sign.

Here is a summary of the key results:

1.\qquad For $\beta\rightarrow\infty$, we may neglect the fourth JJ and reduce
the system to the well-studied flux qubit with three JJs
\cite{Mooij99,Orlando99}. For $\beta\gg1$, there is only a small phase
difference across the fourth junction, with $\theta_{4}=\pm\theta_{4}^{\ast
}=\mathbf{Z}_{\mathrm{flux}}\theta_{4}^{\ast}$ depending on the sign of the
circulating current (i.e., the state of flux qubit, with $\mathbf{Z}%
_{\mathrm{flux}}=\left(  \left\vert 0\right\rangle \left\langle 0\right\vert
-\left\vert 1\right\rangle \left\langle 1\right\vert \right)  _{\mathrm{flux}%
}$). We show that the magnitude of the phase difference can be small
$\theta_{4}^{\ast}\approx\frac{\sqrt{4\alpha^{2}-1}}{2\alpha\beta}\propto
\beta^{-1}$. As shown in Fig.~\ref{fig:FluxQubit}, two SC islands ($c$ and
$d$) are connected by this junction, if we fix the SC phase $\phi_{c}$, then
$\phi_{d}=\varepsilon^{0,1}=\phi_{c}\pm\theta_{4}^{\ast}$ has phase separation
$\Delta\varepsilon=2\theta_{4}^{\ast}\approx\frac{\sqrt{4\alpha^{2}-1}}%
{\alpha\beta}$. The SC island $d$ can be used to coherently control the
coupling between the Majorana fermions of the STIS wire.

2.\qquad There are quantum fluctuations for the phase of the SC island. The
magnitude of quantum fluctuations depends on $\left\{  E_{C,j}\right\}  $ and
$\left\{  E_{J,j}\right\}  $. For $\beta\gg1$, the dynamics associated with
$\theta_{4}$ can be characterized by a harmonic oscillator (HO) Hamiltonian%
\[
H^{\mathrm{HO}}=\frac{p_{\theta_{4}}^{2}}{2M_{4}}+\frac{E_{J,4}}{2}\left(
\theta_{4}-\mathbf{Z}_{\mathrm{flux}}\theta_{4}^{\ast}\right)  ^{2},
\]
where the effective mass is $M_{4}=\frac{1}{8E_{c_{4}}}$ and the canonical
momentum $p_{\theta_{4}}$ satisfies $\left[  \theta_{4},p_{\theta_{4}}\right]
=i$ (with $\hbar\equiv1$). We may rewrite $H^{\mathrm{HO}}=\left(  a^{\dag
}a+1/2\right)  \omega$ and $\theta_{4}=\mathbf{Z}_{\mathrm{flux}}\theta
_{4}^{\ast}+\zeta\left(  a^{\dag}+a\right)  /\sqrt{2}$, where the oscillator
frequency is $\omega=\sqrt{8E_{J}E_{C}}$ and the magnitude of quantum
fluctuations is $\zeta=\left(  \frac{8E_{C}}{E_{J}}\right)  ^{1/4}\beta
^{-1/2}$. We justify that this simple model agrees very well with the general
model characterizing the quantum fluctuations for the flux qubit with coupled JJs.

3.\qquad Various parameters characterizing the flux qubit are also calculated,
including the plasma frequencies $\left\{  \omega_{i}\right\}  _{i=1,2,3}$,
barrier height $\Delta U$, and the tunneling matrix element $t$. For example,
given parameters $\alpha=0.8$ and $\beta=10$, we compute $\left\{  \omega
_{i}\right\}  \approx\left(  2.8,2.3,1.8\right)  \sqrt{E_{J}E_{C}}\sim
\sqrt{E_{J}E_{C}}$, $\Delta U\approx0.26E_{J}$, and $t\approx1.8\sqrt
{E_{J}E_{C}}\exp\left[  -0.7\left(  E_{J}/E_{C}\right)  ^{1/2}\right]  $. We
notice that these parameters for the flux qubit hardly depend on $\beta$ when
$\beta>10$, which verifies the intuition that inserting a large Josephson
junction to the loop has almost no effect to the properties of the flux qubit.

4.\qquad We propose two schemes to implement the SC phase-controller, which
can fix the phase difference between an SC island and a big SC reservoir.

In the following, we provide detailed analysis to justify our design of flux
qubit with four JJs. First, we give the Hamiltonian description for the
system. Then, we calculate the phase separation and quantum fluctuations.
Next, we numerically obtain various quantities such as plasma frequencies,
barrier height, and tunneling matrix element. Our numerical calculation also
verifies our estimates on phase separation and magnitude of quantum
fluctuations. After that, we propose two implementations of the SC
phase-controller. Finally, we derive the energy splitting function $E\left(
\varepsilon\right)  $ that is highly non-linear in terms of $\varepsilon$.

\section{Hamiltonian for Flux Qubit}

The Hamiltonian for a flux qubit consisting of four JJs in series is%
\begin{equation}
H^{\mathrm{flux}}=T+U,
\end{equation}
with the Josephson potential%
\begin{equation}
U=\sum_{j=1,2,3,4}E_{J,j}\left(  1-\cos\theta_{j}\right)  ,
\end{equation}
and the capacitive charging energy%
\begin{equation}
T=\frac{1}{2}\sum_{j=1,2,3,4}C_{j}V_{j}^{2}.
\end{equation}
Here for the $j$-th Josephson junction, $E_{J,j}$ is the Josephson coupling,
$\theta_{j}$ is the gauge-invariant phase difference, $C_{j}$ is the
capacitance, and $V_{j}$ is the voltage across the junction. Suppose that all
the junctions have the same thickness, we have $E_{J,j}$ $\propto A_{j}$ and
$E_{C,j}\equiv\frac{e^{2}}{2C_{j}}\propto A_{j}^{-1}$ where $A_{j}$ is the
area of the junction and $C_{j}$ is the junction capacitance \cite{Tinkham96}.
The quantity $E_{J,j}E_{C,j}=E_{J}E_{C}$ does not depend on $j$.

\begin{figure*}[ptbh]
\begin{center}
\includegraphics[width=16cm]{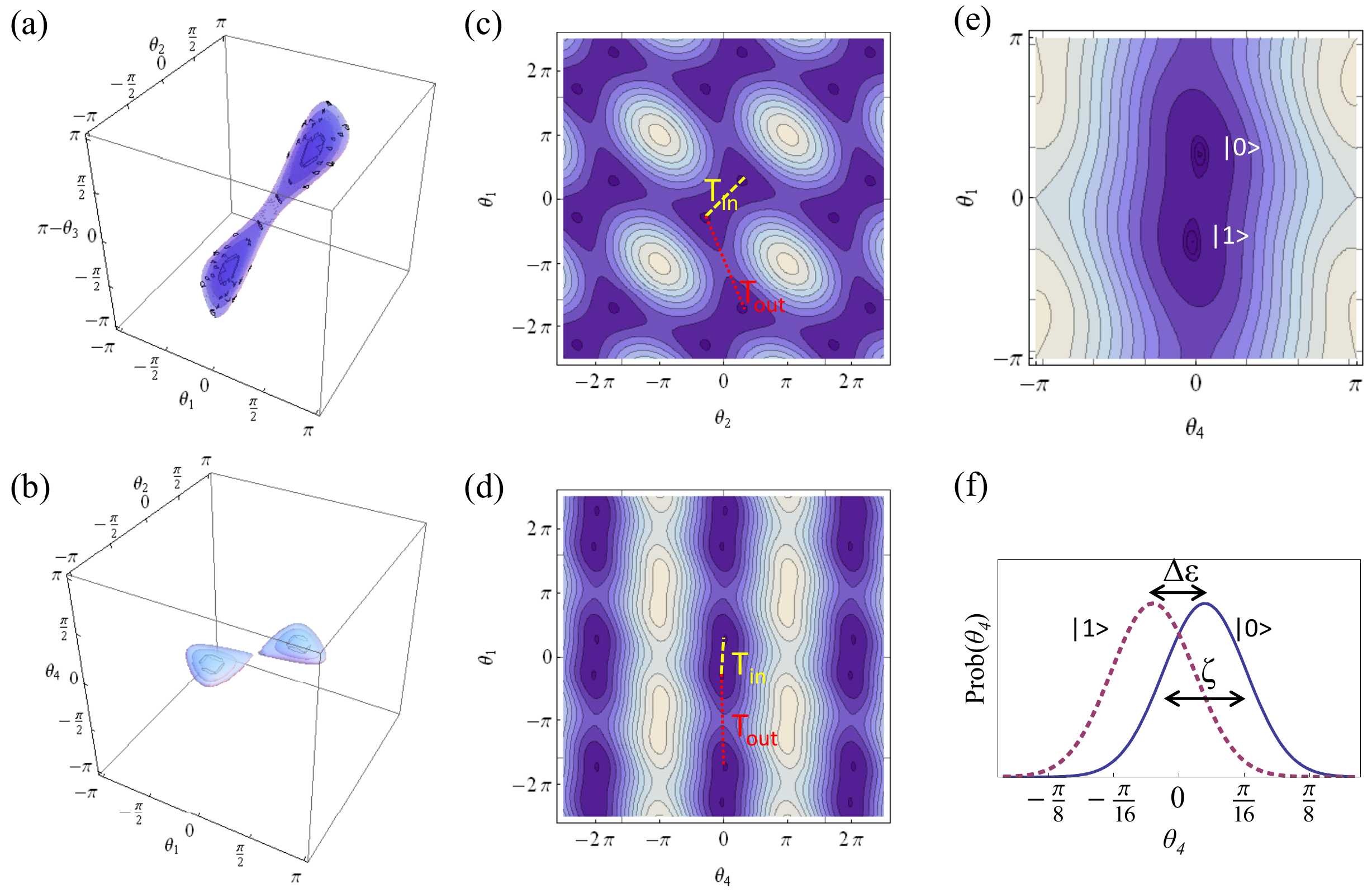}
\end{center}
\caption[fig:Potentialn6]{Contour plots of Josephson energy as a function of
(a) $\left\{  \theta_{1},\theta_{2},\theta_{3}\right\}  $ (with $\theta
_{4}=\pi-\theta_{1}-\theta_{2}-\theta_{3}$). (b) $\left\{  \theta_{1}%
,\theta_{2},\theta_{4}\right\}  $ (with $\theta_{3}=\pi-\theta_{1}-\theta
_{2}-\theta_{4}$). (c) $\left\{  \theta_{1},\theta_{4}\right\}  $ (with
$\theta_{3}=\pi-\theta_{1}-\theta_{2}$ and $\theta_{4}=0$) (d,e) $\left\{
\theta_{1},\theta_{4}\right\}  $ (with $\theta_{1}=\theta_{2}$ and $\theta
_{3}=\pi-2\theta_{1}-\theta_{4}$). (f) Marginal probability distributions of
$\theta_{4}$ associated with states $\left\vert 0\right\rangle _{\mathrm{flux}%
}$ (blue solid line) and $\left\vert 1\right\rangle _{\mathrm{flux}}$ (red
dashed line). The parameters are $E_{J}/E_{C}=80$ and $\left\{  E_{J,i}%
/E_{J}\right\}  _{i=1,2,3,4}=\left\{  1,1,\alpha=0.8,\beta=10\right\}  $.}%
\label{fig:Potentialn6}%
\end{figure*}

The Josephson potential is constraint by the phase relation
\begin{equation}
\sum_{j}\theta_{j}+2f\pi\equiv0\left(  \operatorname{mod}2\pi\right)  ,
\label{eq:PhaseConstraint}%
\end{equation}
This phase relation removes one degree of freedom. We may introduce a 3-vector
$\vec{\theta}=\left(  \theta_{1},\theta_{2},\theta_{3}\right)  $ to
characterize the system with four JJs. (Note that we only need to choose three
independent phases for 3-vector, e.g., $\vec{\theta}=\left(  \theta_{1}%
,\theta_{2},\theta_{4}\right)  $ is also a valid choice.) For $f=1/2$, the
Josephson potential is%
\begin{align}
U  &  =-E_{J,1}\cos\theta_{1}-E_{J,2}\cos\theta_{2}-E_{J,3}\cos\theta_{3}\\
&  +E_{J,4}\cos\left(  \theta_{1}+\theta_{2}+\theta_{3}\right)  .\nonumber
\end{align}
By appropriately choosing the parameter $\left\{  E_{J,j}\right\}  $, there
are only two minima for the Josephson energy at $\pm\left\{  \theta_{j}^{\ast
}\right\}  =\pm\left\{  \theta_{1}^{\ast},\theta_{2}^{\ast},\theta_{3}^{\ast
},\theta_{4}^{\ast}\right\}  $ (see Fig.~\ref{fig:Potentialn6}ab). We may
identify the two levels of the flux qubit as $\left\vert 0\right\rangle
_{\mathrm{flux}}=\left\vert \left\{  \theta_{i}^{\ast}\right\}  \right\rangle
$ and $\left\vert 1\right\rangle _{\mathrm{flux}}=\left\vert \left\{
-\theta_{i}^{\ast}\right\}  \right\rangle $. Thus, we may denoted the two
potential minima as $\left\{  \mathbf{Z}_{\mathrm{flux}}\theta_{j}^{\ast
}\right\}  $, with $\mathbf{Z}_{\mathrm{flux}}=\left(  \left\vert
0\right\rangle \left\langle 0\right\vert -\left\vert 1\right\rangle
\left\langle 1\right\vert \right)  _{\mathrm{flux}}$.

The capacitive charging energy can be regarded as the kinetic energy
associated with the dynamics of $\vec{\theta}$. This is because the voltage
across the junction is given by the Josephson voltage-phase relation
$V_{j}=\left(  \frac{\Phi_{0}}{2\pi}\right)  \dot{\theta}_{j}$
\cite{Tinkham96} and the time derivatives $\left\{  \dot{\theta}_{j}\right\}
$ obey the constraint $\dot{\theta}_{1}+\dot{\theta}_{2}+\dot{\theta}_{3}%
+\dot{\theta}_{4}=0$ (derived from Eq.(\ref{eq:PhaseConstraint})). Thus, we
may write the capacitive charging energy as%
\begin{align}
T  &  =\frac{1}{2}\left(  \frac{\Phi_{0}}{2\pi}\right)  ^{2}\sum
_{i,j=1,2,3}C_{ij}\dot{\theta}_{i}\dot{\theta}_{j}\\
&  =\frac{1}{2}\overrightarrow{\dot{\theta}}^{T}\cdot\mathbf{M}\cdot
\overrightarrow{\dot{\theta}}\\
&  =\frac{1}{2}\vec{p}^{T}\cdot\mathbf{M}^{-1}\cdot\vec{p},
\end{align}
where the capacitive matrix is%
\begin{equation}
C_{ij}=C_{i}\delta_{ij}+C_{4},
\end{equation}
the effective mass tensor is
\begin{equation}
\mathbf{M}=\left(  \frac{\Phi_{0}}{2\pi}\right)  ^{2}\mathbf{C},
\end{equation}
and the canonical momentum is%
\begin{equation}
\vec{p}=\mathbf{M}\cdot\overrightarrow{\dot{\theta}}.
\end{equation}
Therefore, we have reduced the problem to the canonical model of the quantum
system with Hamiltonian
\begin{equation}
H=\frac{1}{2}\vec{p}^{T}\cdot\mathbf{M}^{-1}\cdot\vec{p}+U\left(  \vec{\theta
}\right)  , \label{eq:HamGeneral}%
\end{equation}
where the operators satisfy the commutation relation $\left[  \theta_{j}%
,p_{k}\right]  =i\delta_{jk}$.

Based on this model, we obtain the phase separation and the magnitude of
quantum fluctuations in the next two sections.

\section{Phase Separation between the Potential Minima}

We now calculate the potential minimum $\left\{  \theta_{j}^{\ast}\right\}  $
by introducing a Lagrange variable $\lambda$ associated with the phase
relation (Eq.(\ref{eq:PhaseConstraint})). We study the function%
\begin{equation}
F=-\sum_{j=1,2,3,4}E_{J,j}\cos\theta_{j}-\lambda\left(  \sum_{j}\theta_{j}%
-\pi\right)  .
\end{equation}
The first derivatives all vanish at the extreme point:%
\begin{equation}
E_{J,j}\sin\theta_{j}^{\ast}=\lambda.
\end{equation}
From the phase relation $\sum_{j}\theta_{j}^{\ast}=\sum_{j}\sin^{-1}%
\frac{\lambda}{E_{J,j}}=\pi$, we may solve for $\lambda$.

\begin{figure}[t]
\begin{center}
\includegraphics[width=8cm]{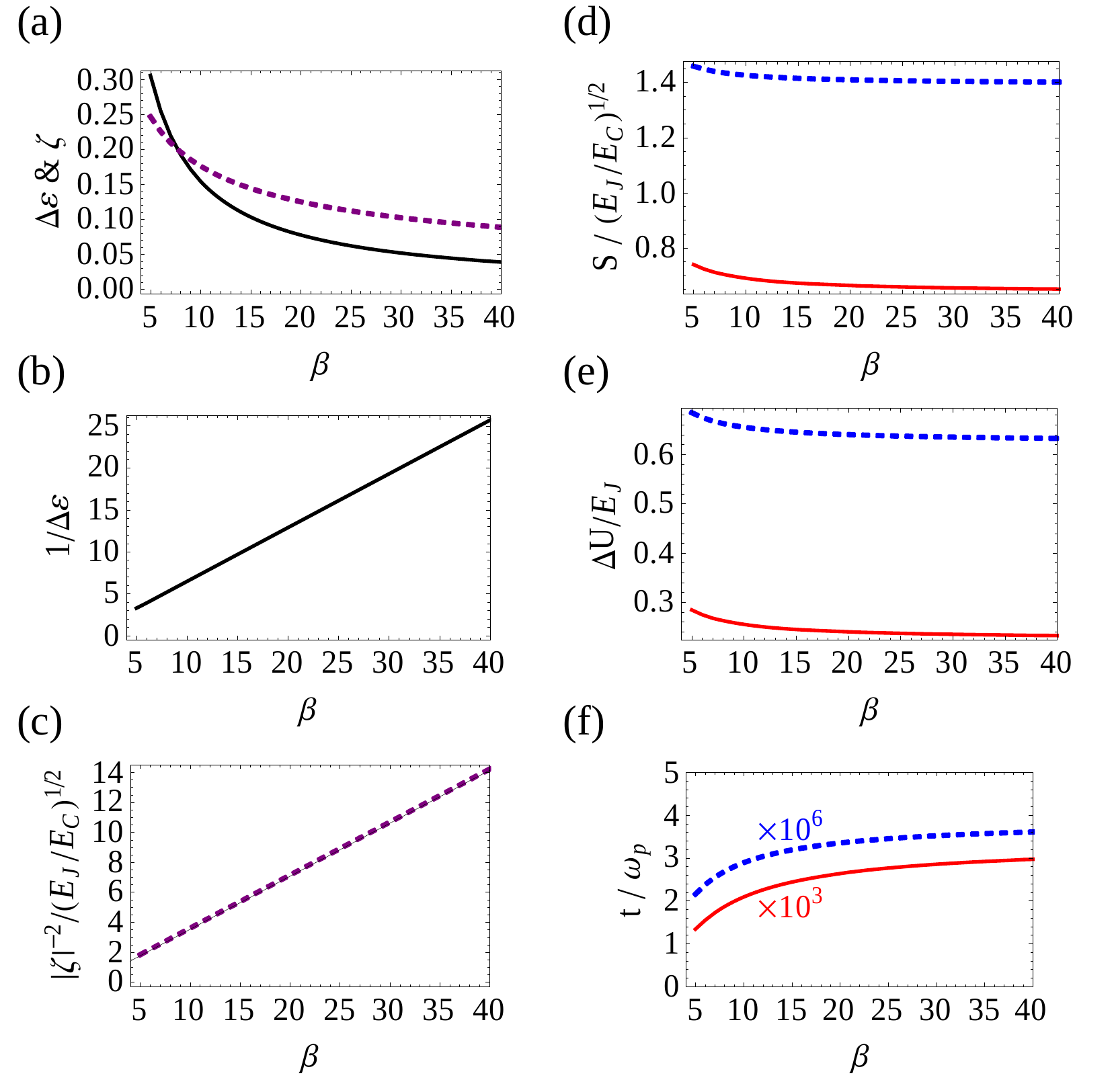}
\end{center}
\caption[fig:BetaDependence2]{Parameters for the flux qubit as a function of
$\beta$. (a,b,c) Phase separation $\Delta\varepsilon$ (dark solid line) and
the magnitude of quantum fluctuations $\zeta$ (purple dashed line) from
numerical calculation. The theoretical prediction from
Eq.~(\ref{eq:QuantumFluctuations}) (gray solid line in (c)) agrees very well
with the numerical calculation. (d,e,f) The two ground states of the flux
qubit can be coupled by intra-cell tunneling (red solid lines) and inter-cell
tunneling (blue dashed lines) \cite{Mooij99,Orlando99}, characterized by the
action $S$, potential barrier $\Delta U$, and tunneling rate $t$. Panels (a)
and (f) assume $E_{J}/E_{C}=80$.}%
\label{fig:BetaDependence2}%
\end{figure}

For our system with four JJs, we can also calculate $\theta_{4}^{\ast}$ by
series expansion with respect to the small parameter $\beta^{-1}$. For
$\beta\rightarrow\infty$, we have the zeroth order expansion $\lambda^{\left(
0\right)  }=\frac{\sqrt{4\alpha^{2}-1}}{2\alpha}$, $\theta_{1}^{\left(
0\right)  }=\theta_{2}^{\left(  0\right)  }=\sin^{-1}\lambda^{\left(
0\right)  }=\cos^{-1}\frac{1}{2\alpha}$, $\theta_{3}^{\left(  0\right)  }%
=\pi-2\cos^{-1}\frac{1}{2\alpha}$ and $\theta_{4}^{\left(  0\right)  }=0$.
Then, to the first order of $\beta^{-1}$, we have $\theta_{4}^{\left(
1\right)  }=\sin^{-1}\frac{\lambda^{\left(  0\right)  }}{\beta}\approx
\frac{\sqrt{4\alpha^{2}-1}}{2\alpha\beta}$. Therefore,\ the phase difference
for the fourth junction is
\begin{equation}
\theta_{4}^{\ast}=\frac{\sqrt{4\alpha^{2}-1}}{2\alpha\beta}+\left(  \beta
^{-2}\right)  .
\end{equation}
As plotted in Fig.~\ref{fig:BetaDependence2}b, the numerically obtained
quantity $\frac{1}{\Delta\varepsilon}=\frac{1}{2\theta_{4}^{\ast}}\propto
\beta$.

\section{Models for Quantum Fluctuations of the SC Phase}

In this section, we consider two models for quantum fluctuations across the
JJs. The quantum fluctuations across the JJs are due to the finite mass matrix
for the Hamiltonian (Eq.(\ref{eq:HamGeneral})), which is proportional to the
capacitance matrix. In the following, we first provide a simple, intuitive
model to characterize the quantum fluctuations associated with $\theta_{4}$
for $\beta\gg1$. Then, we consider the general model of multiple coupled
harmonic oscillators, and obtain the formula for the magnitude of quantum
fluctuations associated with a projected degree of freedom. We find very good
agreement between the two models when $\beta\gg1$, which justifies the simple model.

\subsection{Simple Model -- One Harmonic Oscillator}

Given very large $\beta$, we may neglect the higher order couplings to other
JJs and consider the reduced Hamiltonian for the fourth JJ%
\begin{equation}
H_{J,4}=\frac{1}{2}C_{4}V_{4}^{2}+E_{J,4}\left(  1-\cos\left(  \theta
_{4}-\mathbf{Z}_{\mathrm{flux}}\theta_{4}^{\ast}\right)  \right)  ,
\end{equation}
with displaced potential minimum at $\mathbf{Z}_{\mathrm{flux}}\theta
_{4}^{\ast}$. We then use the harmonic approximation and obtain%
\begin{equation}
H^{\mathrm{HO}}=\frac{p_{\theta_{4}}^{2}}{2M_{4}}+\frac{E_{J,4}}{2}\left(
\theta_{4}-\mathbf{Z}_{\mathrm{flux}}\theta_{4}^{\ast}\right)  ^{2},
\label{eq:Osc2}%
\end{equation}
where $M_{4}=\frac{1}{8E_{C,4}}$ and $E_{C,4}=\frac{e^{2}}{2C_{4}}$. The
oscillator has frequency $\omega=\sqrt{8E_{J,4}E_{C,4}}=\sqrt{8E_{J}E_{C}}$
and characteristic length
\begin{equation}
\zeta=\left(  \frac{8E_{C,4}}{E_{J,4}}\right)  ^{1/4}=\left(  \frac{8E_{C}%
}{E_{J}}\right)  ^{1/4}\beta^{-1/2}. \label{eq:QuantumFluctuations}%
\end{equation}
Here $\zeta$ is also the magnitude of quantum fluctuations.

\subsection{General Model -- Coupled Harmonic Oscillators}

We now consider the general model of multiple coupled harmonic oscillators,
and obtain the formula for the magnitude of quantum fluctuations associated
with a projected degree of freedom.

Close to the potential minimum $\left\{  \mathbf{Z}_{\mathrm{flux}}\theta
_{j}^{\ast}\right\}  $, we may expand the potential function to the second
order of $\vec{x}\equiv\vec{\theta}-\mathbf{Z}_{\mathrm{flux}}\vec{\theta
}^{\ast}$%
\begin{equation}
U=U_{\min}+\frac{1}{2}\sum_{i,j=1,2,3}K_{ij}x_{i}x_{j}+O\left(  x^{3}\right)
\end{equation}
with%
\begin{equation}
K_{ij}=\left.  \frac{d^{2}U}{d\theta_{i}d\theta_{j}}\right\vert _{\vec{\theta
}=\vec{\theta}^{\ast}}.
\end{equation}
The effective Hamiltonian around the minimum describes a system of coupled
Harmonic oscillators:%
\begin{align}
H_{\mathrm{oscillator}}  &  =\frac{1}{2}\sum_{i,j=1,2,3}M_{ij}\dot{x}_{i}%
\dot{x}_{j}+\frac{1}{2}\sum_{i,j=1,2,3}K_{ij}x_{i}x_{j}\\
&  =\frac{1}{2}\vec{p}^{T}\cdot\mathbf{M}^{-1}\cdot\vec{p}+\frac{1}{2}\vec
{x}^{T}\cdot\mathbf{K}\cdot\vec{x}%
\end{align}
where we have used the definition $\vec{p}=\mathbf{M}\cdot\overrightarrow
{\dot{\theta}}=\mathbf{M}\cdot\overrightarrow{\dot{x}}$.

To solve for the system of coupled oscillators, we perform the following
transformation to make the mass matrix/tensor isotropic. For real and
symmetric mass matrix ($M_{ij}=M_{ji}$), there is an orthogonal transformation
$V_{1}$ (with $V_{1}^{-1}=V_{1}^{T}$) that diagonalizes the mass matrix%
\begin{equation}
V_{1}\mathbf{M}V_{1}^{T}=\mathbf{\Lambda}%
\end{equation}
where $\Lambda_{ij}=\lambda_{i}\delta_{ij}$ is the diagonal matrix. The
eigenvalue $\lambda_{i}$ is effective mass along the $i$-th principle axis. By
inverting both sides, we have $V_{1}\mathbf{M}^{-1}V_{1}^{T}=\mathbf{\Lambda
}^{-1}$. For diagonal matrix, we may also define $\left(  \Lambda^{\pm
1/2}\right)  _{ij}=\lambda_{i}^{\pm1/2}\delta_{ij}$. Introducing the
transformation%
\begin{align}
\vec{x}^{\prime}  &  =\Lambda^{1/2}V_{1}\cdot\vec{x}\\
\vec{p}^{\prime}  &  =\Lambda^{-1/2}V_{1}\cdot\vec{p}%
\end{align}
we have%
\begin{equation}
H_{\mathrm{oscillator}}=\frac{1}{2}\vec{p}^{\prime T}\cdot\vec{p}+\frac{1}%
{2}\vec{x}^{\prime T}\cdot\mathbf{\tilde{K}}\cdot\vec{x}%
\end{equation}
where%
\begin{equation}
\mathbf{\tilde{K}=}\Lambda^{-1/2}\left(  V_{1}KV_{1}^{T}\right)
\Lambda^{-1/2}.
\end{equation}
Finally, we diagonalize the real symmetric matrix $\mathbf{\tilde{K}}$ via
orthogonal transformation $V_{2}$%
\begin{equation}
V_{2}\mathbf{\tilde{K}}V_{2}^{T}=\mathbf{\Omega}%
\end{equation}
where $\Omega_{ij}=\omega_{i}^{2}\delta_{ij}$ is the diagonal matrix. Overall,
the position and momentum transform as
\begin{subequations}
\label{eq:Transformation}%
\begin{align}
\vec{y}  &  =V_{2}\vec{x}^{\prime}=V_{2}\Lambda^{1/2}V_{1}\cdot\vec{x}\\
\vec{q}  &  =V_{2}\vec{p}^{\prime}=V_{2}\Lambda^{-1/2}V_{1}\cdot\vec{p}%
\end{align}
The eigenvalue $\omega_{i}^{2}$ is the square of the $i$-th oscillator
frequency (also called plasma frequency). Given parameters $\alpha=0.8$ and
$\beta=10$, we calculate the plasma frequencies $\left\{  \omega_{i}\right\}
_{i=1,2,3}=\left\{  2.8,2.3,1.8\right\}  \sqrt{E_{J}E_{C}}$. We may also vary
the parameter $\beta$, and observe that the plasma frequencies only depend
very weakly for $\beta\gg1$.

Note that position and momentum have different transformations $V_{2}%
\Lambda^{1/2}V_{1}$ and $V_{2}\Lambda^{-1/2}V_{1}$. Furthermore, these
transformations are not orthogonal transformations. However, as long as the
transformations preserve the commutation relation $\left[  \hat{y}_{j},\hat
{q}_{k}\right]  =\left[  \hat{x}_{j},\hat{p}_{k}\right]  =i\delta_{jk}$, we
can still perform quantization over the transformed coordinate.

Following this procedure, we can numerically compute the magnitude of quantum
fluctuations of $\theta_{4}$ as detailed below.

\subsection{Quantum Fluctuations in SC Phase}

We now quantize the phase difference across the fourth junction $\theta_{4}$.
Near the potential minimum at $\left\{  \theta_{i}^{\ast}\right\}  $, we
perform the transformation of Eq.\ (\ref{eq:Transformation}) along with the
quantization $\hat{y}_{i}=\frac{\hat{a}_{i}^{\dag}+\hat{a}_{i}}{\sqrt{2}}$ and
$\hat{q}_{i}=\frac{\hat{a}_{i}^{\dag}-\hat{a}_{i}}{\sqrt{2}i}$, we obtain the
Hamiltonian for three uncoupled harmonic oscillators%
\end{subequations}
\begin{equation}
\tilde{H}_{\mathrm{oscillator}}=\sum_{i=1,2,3}\hbar\omega_{i}\left(  \hat
{a}_{i}^{\dag}\hat{a}_{i}+\frac{1}{2}\right)  ,
\end{equation}
with eigenfrequencies of $\left\{  \omega_{i}\right\}  _{i=1,2,3}$. Each
oscillatory mode may induce quantum fluctuations in $\theta_{4}$, with
characteristic length scale $\zeta_{i}$. The operator form of $\theta_{4}$ can
be written as%
\begin{equation}
\theta_{4}=\mathbf{Z}_{\mathrm{flux}}\theta_{4}^{\ast}+\sum_{i=1,2,3}\zeta
_{i}\frac{\hat{a}_{i}^{\dag}+\hat{a}_{i}}{\sqrt{2}}%
\end{equation}

We calculate the values of $\zeta_{i}$ as the following. In the $y$%
-coordinate, the characteristic displacement vector for the $i$-th mode is
$\vec{l}_{i}^{\left(  y\right)  }=\sqrt{\frac{\hbar}{\omega_{i}}}\vec{e}%
_{i}^{\left(  y\right)  }$, with unit vector $\vec{e}_{i}$ along the $i$-th
direction. We may transform this back to the $x$-coordinate, $\vec{l}%
_{i}^{\left(  x\right)  }=\left(  V_{2}\Lambda^{1/2}V_{1}\right)  ^{-1}%
\cdot\sqrt{\frac{\hbar}{\omega_{i}}}\vec{e}_{i}^{\left(  y\right)  }$. Note
that $\vec{l}_{i}^{\left(  x\right)  }$ is no longer orthogonal. From $\vec
{l}_{i}^{\left(  x\right)  }$, we can obtain the characteristic fluctuating
scale of $\zeta_{i}=\left\vert \sum_{k=1,2,3}\left(  \vec{l}_{i}^{\left(
x\right)  }\right)  _{k}\right\vert $. The magnitude of quantum fluctuations
of $\theta_{4}$ can be computed as
\begin{equation}
\zeta=\left(  \sum_{i=1,2,3}\left\vert \zeta_{i}^{2}\right\vert \right)
^{1/2} \label{eq:QuantumFluctuations2}%
\end{equation}
for independent fluctuations from the three uncoupled harmonic oscillators.

As plotted in Fig.~\ref{fig:BetaDependence2}c, the numerically obtained value
for $\zeta$ (using Eq.~(\ref{eq:QuantumFluctuations2})) agrees very well with
the prediction from the simple model (using Eq.~(\ref{eq:QuantumFluctuations}%
)), which scales as $\beta^{-1/2}$. Therefore, the simple model of
Eq.~(\ref{eq:QuantumFluctuations}) provides a reliable description to
characterize the dynamics associated with $\theta_{4}$.

\section{Tunneling Matrix Element}

\subsection{WKB method}

We use the WKB method to estimate the tunneling matrix element
\cite{Orlando99}. The action associated with the pathway $\vec{\theta}\left(
r\right)  $ from $\vec{\theta}\left(  0\right)  =\vec{\theta}_{a}$ to
$\vec{\theta}\left(  1\right)  =\vec{\theta}_{b}$ is%
\begin{equation}
S=\int_{\vec{\theta}_{a}}^{\vec{\theta}_{b}}\sqrt{2\left(  U-E\right)  }%
\sqrt{d\vec{\theta}^{T}\cdot\mathbf{M}\cdot d\vec{\theta}},
\end{equation}
and the tunneling matrix element can be estimated as
\begin{equation}
t\approx\frac{\hbar\omega}{2\pi}~e^{-S/\hbar}.
\end{equation}
The phase space $\vec{\theta}$ has period $2\pi~$for all three directions. We
may introduce the unit cell with volume $\left(  2\pi\right)  ^{3}$ and three
basis vectors $\vec{a}_{1}=2\pi\left(  1,0,0\right)  $, $\vec{a}_{2}%
=2\pi\left(  0,1,0\right)  $, and $\vec{a}_{3}=2\pi\left(  0,0,1\right)  $.
(Note that we may choose the shape of the unit cell for our convenience.)
Regarding the tunneling pathway, we may choose the initial point $\vec{\theta
}_{a}^{\ast}=-\vec{\theta}^{\ast}$, while the choice for final point is not
unique as $\vec{\theta}_{b}=\vec{\theta}^{\ast}+\sum_{i}n_{i}\vec{a}_{i}$ for
integers $\left\{  n_{i}\right\}  $. However, we may require that the final
point, $\vec{\theta}_{b}^{\ast}$, be the one that has the minimum action from
the initial point, and we choose the unit cell so that it includes the minimum
action pathway that connects $\vec{\theta}_{a}^{\ast}$ and $\vec{\theta}%
_{b}^{\ast}$. After this procedure, the intra-cell tunneling has the minimum
action, compared to all inter-cell tunneling pathways.

\subsection{Pathway with minimum action}

We should use the pathway $\vec{\theta}\left(  r\right)  $ with extreme action
for the WKB method, i.e.,%
\[
\frac{\delta S}{\delta\vec{\theta}\left(  r\right)  }=0.
\]
We obtain these extreme pathways using the following approach. First, we
discretize the integration%
\begin{equation}
S=\sum_{i=1}^{N}\left\vert \vec{x}_{i}-\vec{x}_{i-1}\right\vert ~f\left(
\frac{\vec{x}_{i}+\vec{x}_{i-1}}{2}\right)  ,
\end{equation}
with $\vec{x}_{0}=\vec{\theta}_{a}$ and $\vec{x}_{N}=\vec{\theta}_{b}$. Then
we calculate the derivatives with respect to $\vec{r}_{i}$%
\begin{align}
\frac{\delta S}{\delta\vec{r}_{i}} &  =\frac{\Delta\vec{x}_{i}}{\left\vert
\Delta\vec{x}_{i}\right\vert }f\left(  \frac{\vec{x}_{i}+\vec{x}_{i-1}}%
{2}\right)  -\frac{\Delta\vec{x}_{i+1}}{\left\vert \Delta\vec{x}%
_{i+1}\right\vert }f\left(  \frac{\vec{x}_{i+1}+\vec{x}_{i}}{2}\right)
\nonumber\\
&  +\left\vert \Delta\vec{x}_{i}\right\vert \bigtriangledown f\left(
\frac{\vec{x}_{i}+\vec{x}_{i-1}}{2}\right)  +\left\vert \Delta\vec{x}%
_{i+1}\right\vert \bigtriangledown f\left(  \frac{\vec{x}_{i+1}+\vec{x}_{i}%
}{2}\right)  .
\end{align}
For extreme pathway, these derivatives should vanish. If we start with an
non-extreme pathway with non-vanishing derivatives, we can update the pathway
so that the updated pathway becomes closer to the extreme pathway. By
repeating the update procedure many times, we will obtain a pathway that is
very close to the extreme pathway. Since we know in advance that we are
looking for the pathway that gives the minimum action, we apply the following
update rules for the $k$th update:%
\begin{equation}
\vec{r}_{i}^{\left(  k+1\right)  }=\vec{r}_{i}^{\left(  k\right)  }%
-\epsilon\frac{\delta S}{\delta\vec{r}_{i}}%
\end{equation}
where $\epsilon$ determines the evolution rate and $i=1,2,\cdots,N-1$. For
sufficiently small evolution rate%
\begin{equation}
S\left[  \left\{  \vec{r}_{i}^{\left(  k+1\right)  }\right\}  \right]
-S\left[  \left\{  \vec{r}_{i}^{\left(  k\right)  }\right\}  \right]
=-\epsilon\left(  \frac{\delta S}{\delta\vec{r}_{i}}\right)  ^{2}+O\left(
\epsilon^{2}\right)  \leq0,
\end{equation}
which ensures continuous reduction of the action.

\subsection{Tunneling rates and $\beta$ Dependence}

This algorithm gives us the correct pathway that locally minimize the action
between neighboring potential wells. We find that the pathways are essentially
the straight lines connecting the different minima, with no significant
difference in terms of the action values. For example, given parameters
$\alpha=0.8$ and $\beta=10$, the intra-cell action is $S_{in}\approx
0.7\hbar\sqrt{E_{J}/E_{C}}$ and the smallest inter-cell action is
$S_{out}\approx1.4\hbar\sqrt{E_{J}/E_{C}}$. For $E_{J}/E_{C}\approx80$, we
will have $t_{2}/t_{1}\approx\exp\left[  \left(  S_{1}-S_{2}\right)
/\hbar\right]  \sim10^{-3}\ll1$. The barrier height for intra-cell tunneling
is $\Delta U=0.25E_{J}$.

When $\beta\rightarrow\infty$, all quantities (potential minimum position,
plasma frequencies, action for tunneling, energy barrier, and tunneling matrix
element) reduce to the case with three JJs. As illustrated in
Fig.~\ref{fig:BetaDependence2}, the deviation scales as $\beta^{-1}$. For
$\beta\geq10$, the perturbation from $\beta$ is very small.

For practical parameters of mesoscopic aluminum junctions with critical
current density $500$ A/cm$^{2}$ \cite{Mooij99,Orlando99}, a junction with an
area of $A=0.2\times0.4$ $\mu m^{2}$ can achieve $E_{J}\approx200$ GHz and
$E_{J}/E_{C}\approx80$, which corresponds to the first two junctions
$E_{J,1}=E_{J,2}=E_{J}$. Since $E_{J,j}\propto A_{j}$, the fourth junction
$E_{J,4}=\beta E_{J}$ should have an area of approximately $1\times1$ $\mu
m^{2}$ to achieve $\beta\approx10.$

\section{Phase-Controllers}

For the STIS quantum wire, we would like to fix the phase difference between
two disconnected SC islands. For example, we would like have $\phi_{l}%
-\phi_{u}=\pi/2$, $\phi_{r}-\phi_{u}=\pi/2$, $\phi_{c}-\phi_{u}=\phi_{c}+\pi$,
as shown in Fig.~\ref{fig:TopoFluxQubits}ac. The idea is to connect the two SC
islands via a phase-controller. The phase-controller has two large SC islands
with a controllable phase difference $\gamma$. Using large SC islands for
phase-controllers reduces quantum fluctuations in the SC phase.

In this section, we consider two approaches to building a phase-controller
using Josephson junctions (JJs). The phase difference $\gamma$ between the two
large SC islands can be controlled by either the external magnetic flux
$\gamma=\gamma\left(  \Phi_{x}\right)  $ or the electric current
$\gamma=\gamma\left(  I\right)  $, as detailed below.

\begin{figure}[tbh]
\begin{center}
\includegraphics[width=8.7cm]{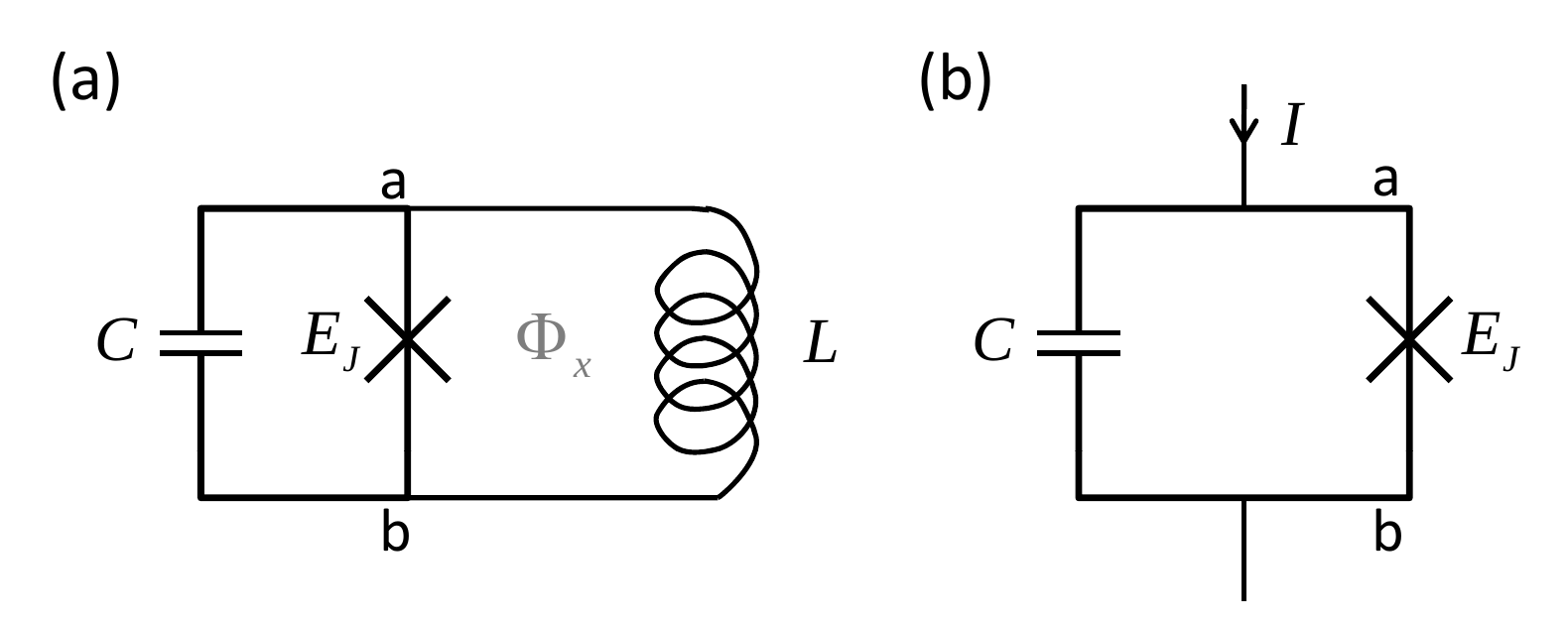}
\end{center}
\caption[fig:PhaseController]{The phase-controllers can establish desired
phase difference $\gamma$ between SC islands $a$ and $b$. For example, (a) the
flux phase-controller with external magnetic flux $\Phi_{x}$, and (b) the
current phase-controller with external electric current $I$.}%
\label{fig:PhaseController}%
\end{figure}

\subsection{Flux phase-controller}

The \emph{flux phase-controller} uses an rf SQUID loop that is interrupted by
a single Josephson junction (Fig.~\ref{fig:PhaseController}a). We may change
the external magnetic flux enclosed by the loop, to induce the desired phase
difference between the two SC\ islands, $a$ and $b$. The rf SQUID loop has
inductance $L$, and the JJ has Josephson coupling energy $E_{J}$ and
capacitive charging energy $E_{C}$ ($=e^{2}/2C$). The Hamiltonian for the flux
phase-controller is%
\begin{equation}
H=U+T,
\end{equation}
where the potential energy is%
\begin{equation}
U=\frac{1}{2L}\left(  \Phi_{x}-\frac{\Phi_{0}}{2\pi}\gamma\right)  ^{2}%
+E_{J}\left(  1-\cos\gamma\right)
\end{equation}
with $\Phi_{x}$ for the external flux enclosed by the loop and $\gamma$ for
the gauge invariant phase difference across the junction, and the capacitive
charging energy is%
\begin{equation}
T=\frac{1}{2}CV^{2}%
\end{equation}
with the voltage $V=\frac{\Phi_{0}}{2\pi}\frac{d\gamma}{dt}$ and the effective
mass $m_{eff}=C\left(  \frac{\Phi_{0}}{2\pi}\right)  ^{2}.$

We find that the potential minimum satisfies the condition%
\begin{equation}
\Phi_{x}=\frac{\Phi_{0}}{2\pi}\gamma+E_{J}L\frac{2\pi}{\Phi_{0}}\sin\gamma,
\end{equation}
This expression can be used to determine $\gamma$ as a function of $\Phi_{x}$.
For $L\rightarrow0$, we have $\gamma=-2\pi\frac{\Phi_{x}}{\Phi_{0}}$. For
sufficiently small $L$ (satisfying $L<\frac{\Phi_{0}^{2}}{4\pi^{2}E_{J}}$),
this relation is still single valued for all $\gamma$. Therefore, we can
deterministically control the gauge invariant phase difference $\gamma$ by
applying an appropriate $\Phi_{x}$.

We then consider the perturbation around the potential minimum and obtain the
plasma frequency $\omega_{p}\approx\left(  LC\right)  ^{-1/2}$ for $L\ll
\frac{\Phi_{0}^{2}}{4\pi^{2}E_{J}}$. The quantum fluctuations of $\gamma$
around the potential minimum has characteristic scale
\begin{equation}
\zeta_{\mathrm{flux}}=\sqrt{\frac{\hbar}{m_{eff}\omega_{p}}}\approx2\sqrt{\pi
}\left(  \frac{E_{c}}{E_{L}}\right)  ^{1/4}.
\end{equation}

\subsection{Current phase-controller}

Alternatively, we may also control the phase differences via the external
current $I$ through a Josephson junction (Fig.~\ref{fig:PhaseController}a).
The Hamiltonian for such a \emph{current phase-controller} is%
\begin{equation}
H=U+T,
\end{equation}
where the potential energy is%
\begin{equation}
U=-I\frac{\Phi_{0}}{2\pi}\gamma+E_{J}\left(  1-\cos\gamma\right)
\end{equation}
with $\gamma$ for the gauge invariant phase difference across the junction,
and the capacitive charging energy is%
\begin{equation}
T=\frac{1}{2}CV^{2}%
\end{equation}
with the voltage $V=\frac{\Phi_{0}}{2\pi}\frac{d\gamma}{dt}$ and the effective
mass $m_{eff}=C\left(  \frac{\Phi_{0}}{2\pi}\right)  ^{2}$.

As long as $I<I_{c}\equiv\frac{2\pi}{\Phi_{0}}E_{J}$, we have the potential
minimum at
\begin{equation}
\gamma=\sin^{-1}I/I_{c}.
\end{equation}
We then consider the perturbation around the potential minimum and obtain the
plasma frequency $\omega_{p}\approx\left(  \frac{2eI_{c}\cos\gamma^{\ast}%
}{\hbar C}\right)  ^{1/2}$. The quantum fluctuations of $\gamma$ around the
potential minimum has characteristic scale
\begin{equation}
\zeta_{\mathrm{current}}=\sqrt{\frac{\hbar}{m_{eff}\omega_{p}}}=\left(
\frac{8}{\cos\gamma^{\ast}}\frac{E_{c}}{E_{J}}\right)  ^{1/4}.
\end{equation}

\subsection{Comparison between flux and current phase-controllers}

Both flux and current phase-controllers enable us to reliably induce phase
difference between two SC islands. We compare the two phase-controllers in the
following aspects:

(1) Tunable phase rage: A single flux phase-controller can create all desired
phase differences in the range $\left[  -\pi,\pi\right]  $. In contrast, a
single current phase-controller can only create phase difference in the range
$\left(  -\pi/2,\pi/2\right)  $. However, by using two or more current
phase-controllers in series, it is possible to create arbitrary phase differences.

(2) Quantum fluctuations: Both controllers have similar scaling for the
quantum fluctuations, which scales as $\left(  E_{c}/E_{L,J}\right)  ^{1/4}$.
In order to minimize the quantum fluctuations, we may use the JJ loops with a
small inductance (i.e., $E_{L}>E_{J}$) for flux phase-controller, while we may
use the JJ loops with large Josephson coupling energy $E_{J}$ for current
phase-controller. Note that for current phase-controller the quantum
fluctuations becomes unfavorably large when $\gamma\approx\pi/2$, which can be
overcome by using two or more controllers in series to reduce the fluctuations.

\subsection{Parameters}

We may estimate the quantum fluctuations for practical devices. According to
the experimental parameters of the large Josephson junctions \cite{Martinis02}%
: the charging energy $E_{c}\equiv e^{2}/2C=0.15$ mK and the Josephson
coupling energy $E_{J}=I_{c}\Phi_{0}/2\pi=500$ K, which gives us
$\zeta_{\mathrm{current}}\approx(8\ast\frac{0.00015}{500})^{1/4}=0.04$. It is
also possible to build a SQUID loop with very small inductive energy
$E_{L}\equiv\Phi_{0}^{2}/2L=645$ K \cite{Friedman00}, and we can obtain
$\zeta_{\mathrm{flux}}\approx2\sqrt{\pi}\ast(\frac{0.00015}{645})^{1/4}%
\approx0.08$. By further increasing the junction area, we may further decrease
$E_{c}$ and increase $E_{J}$, which should give us more reduced quantum
fluctuations from the phase-controller.

\section{Brief Derivation for Energy Splitting $E\left(  \varepsilon\right)
$}

We now briefly derive the energy splitting function $E\left(  \varepsilon
\right)  $, which is a highly non-linear function of $\varepsilon$. The
derivation mostly follows Ref.~\cite{FuL08}.

We start with the effective\ Hamiltonian \cite{FuL08}%
\begin{equation}
H^{\mathrm{STIS}}=-iv_{F}\tau^{x}\partial_{x}+\delta_{\varepsilon}\tau
^{z}\mathrm{,\ }%
\end{equation}
with $\delta_{\varepsilon}=-\Delta_{0}\sin\varepsilon/2$ (differed from
\cite{FuL08} by a minus sign, due to a slightly different assignment of SC
phases). The Hamiltonian can be written as%
\begin{equation}
H^{\mathrm{STIS}}\left(  k\right)  =v_{F}k\tau^{x}+\delta_{\varepsilon}%
\tau^{z}=\left(
\begin{array}
[c]{cc}%
\delta_{\varepsilon} & v_{F}k\\
v_{F}k & -\delta_{\varepsilon}%
\end{array}
\right)  ,
\end{equation}
where $k$ is the wave vector in the quantum wire. The eigen-energies for
$H^{\mathrm{STIS}}\left(  k\right)  $ are $E^{\pm}\left(  \delta_{\varepsilon
},k\right)  =\pm\sqrt{\delta_{\varepsilon}^{2}+v_{F}^{2}k^{2}}$. For a finite
STIS quantum wire with length $L$, it can only support a discretized set of
wave vectors satisfying the boundary condition \cite{FuL08}%
\begin{equation}
\tan kL=-\frac{v_{F}k}{\delta_{\varepsilon}}=kL/\Lambda_{\varepsilon},
\label{eq:BC}%
\end{equation}
with a dimensionless parameter
\begin{equation}
\Lambda_{\varepsilon}\equiv-\delta_{\varepsilon}L/v_{F}.
\end{equation}
For given $\Lambda_{\varepsilon}$, we may solve Eq. (\ref{eq:BC}) and obtain a
set of solutions $kL=f_{n}\left(  \Lambda_{\varepsilon}\right)  $ with index
$n=0,1,2,\cdots$ for different bands. The function $f_{n}\left(  y\right)  $
is just the inverse function of $y=x/\tan\left(  x\right)  $ associated with
the $n$th invertible domain.

For the lowest band, we have
\begin{equation}
E\left(  \varepsilon\right)  \equiv E^{+}\left(  \delta_{\varepsilon}%
,k_{0}\right)  =\Delta E\sqrt{\Lambda_{\varepsilon}^{2}+f_{0}^{2}\left(
\Lambda_{\varepsilon}\right)  },
\end{equation}
with $\Delta E=v_{F}/L$. Note that $kL=f_{0}\left(  \Lambda_{\varepsilon
}\right)  $ is purely imaginary for $\Lambda_{\varepsilon}\in\left(
1,\infty\right)  $, and it is real for $\Lambda_{\varepsilon}\in\left(
-\infty,1\right)  $. Physically, imaginary $kL$ corresponds to localized MFs
at the ends of the quantum wire, and real $kL$ indicates delocalized MFs.
Those higher bands (with $n\geq1$) are associated with the excitation modes of
the quantum wire, with excitation energy at least $\Delta E$ \cite{FuL08}.

In summary, we have%
\begin{equation}
\frac{E\left(  \varepsilon\right)  }{\Delta E}=\sqrt{\Lambda_{\varepsilon}%
^{2}+f_{0}^{2}\left(  \Lambda_{\varepsilon}\right)  },
\end{equation}
which is plotted in Fig~\ref{fig:Potential}a.

\end{document}